\RequirePackage{fix-cm}
\documentclass[smallextended]{svjour3}       
\smartqed  
\usepackage{graphicx}
\usepackage{color}
\begin{document}

\title{Variable-coefficient symbolic computation approach for finding multiple rogue wave
solutions of nonlinear system with variable
coefficients\thanks{Project supported by National Natural Science
Foundation of China (Grant No 81960715).}}


\author{ Jian-Guo Liu$^{*}$, Wen-Hui Zhu, Yan He
}


\institute{Jian-Guo Liu(*Corresponding author), Yan He\at
              College of Computer, Jiangxi University of
Traditional Chinese Medicine, Jiangxi 330004, China\\
              \email{20101059@jxutcm.edu.cn}\\
\and Jian-Guo Liu\at
 School of science, Beijing University of Posts
and Telecommunications, Beijing 100876, China\\
\and  Wen-Hui Zhu \at
Institute of artificial intelligence, Nanchang
Institute of Science and Technology, Jiangxi 330108, China\\
              \email{415422402@qq.com}}

\date{Received: date / Accepted: date}

\maketitle

\begin{abstract}
In this paper, a variable-coefficient symbolic computation approach
 is proposed to solve the multiple rogue wave solutions of
nonlinear equation with variable coefficients. As an application, a
(2+1)-dimensional variable-coefficient Kadomtsev-Petviashvili
equation is investigated. The multiple rogue wave solutions  are
obtained and their dynamics features are shown in some 3D and
contour plots.

\keywords{variable-coefficient symbolic computation approach, rogue
wave, variable coefficient Kadomtsev-Petviashvili equation.}
\subclass{35C08 \and 68M07 \and 33F10}
\end{abstract}

\section{Introduction}
\label{intro} \quad In this paper, the following (2+1)-dimensional
variable-coefficient Kadomtsev-Petviashvili (vcKP) equation is
investigated [1]
\begin{eqnarray}
\alpha (t) u_x^2+\alpha (t) u u_{xx}+\beta (t)
   u_{xxxx}-\gamma (t) u_{yy}+u_{xt}=0,
\end{eqnarray}where $u=u(x,y,t)$ describes amplitude of the long wave of two-dimensional
fluid domain on varying topography or in turbulent ow over a sloping
bottom. $\alpha(t)$, $\beta(t)$ and $\gamma(t)$ are arbitrary real
functions. The solitonic solution [1],  Wronskian and Gramian
solutions [2], B\"{a}cklund transformation [3], breather wave
solutions [4], lump and interactions solutions [5, 6] of Eq. (1)
have been studied.

\quad Rogue wave has important applications in ocean's waves [7],
optical fibers [8], Bose-Einstein condensates [9] and other fields.
Rogue wave solutions of many integrable equations have been
investigated [10-14]. Recently, a symbolic computation approach to
obtain the multiple rogue wave solutions is proposed by Zhaqilao
[15]. But the main application of this method is
constant-coefficient integrable equation [16-18], which is not
suitable for variable-coefficient integrable equation. So, we give
an improved method named variable-coefficient symbolic computation
approach (vcsca) to solve this problem and apply this method to Eq.
(1), which will be the main work of our paper.

\quad The organization of this paper is as follows. Section 2
proposes a vcsca; Section 3 gives the 1-rogue wave solutions;
Section 4 obtains the 3-rogue wave solutions; Section 5 presents the
6-rogue wave solutions; Section 6 gives this conclusions.

\section{Modified symbolic computation approach} \label{sec:2}
\quad Here, we present a vcsca to find the multiple rogue wave
solutions of variable
coefficient integrable equation\\

Step1. Instead of the traveling wave transformation in Ref. [15], we
make a non-traveling wave transformation  $\upsilon=x-\omega(t)$ in
the following nonlinear system with variable coefficients
{\begin{eqnarray}\Xi(u, u_t, u_x, u_y, u_{xy},
\cdots)=0,\end{eqnarray} }Eq. (2) is reduced to a (1+1)-dimensional
equation{\begin{eqnarray}\Xi(u, u_\upsilon, u_y, u_{\upsilon y},
\cdots)=0.
\end{eqnarray} }
Step2. By Painlev\'{e} analysis, we make the following
transformation {\begin{eqnarray}u(\upsilon,
y)=\frac{\partial^n}{\partial\upsilon^m}ln\xi(\upsilon, y).
\end{eqnarray} }$m$ can be derived by balancing the order of the highest
derivative term and nonlinear term.

Step3. Assuming {\begin{eqnarray}\xi(\upsilon, y)=F_{n+1}(\upsilon,
y)+2 \nu y P_n(\upsilon, y)+2 \mu \upsilon Q_n(\upsilon,
y)+(\mu^2+\nu^2) F_{n-1}(\upsilon, y),
\end{eqnarray} }with
{\begin{eqnarray}F_n(\upsilon, y)&=&\sum^{n(n+1)/2}_{k=0}\sum^{k}_{i=0}a_{n(n+1)-2k,2i}y^{2i}\upsilon^{n(n+1)-2k},\nonumber\\
 P_n(\upsilon, y)&=&\sum^{n(n+1)/2}_{k=0}\sum^{k}_{i=0}b_{n(n+1)-2k,2i}\upsilon^{2i}y^{n(n+1)-2k},\nonumber\\
  Q_n(\upsilon, y)&=&\sum^{n(n+1)/2}_{k=0}\sum^{k}_{i=0}c_{n(n+1)-2k,2i}y^{2i}\upsilon^{n(n+1)-2k},\nonumber
\end{eqnarray} }$F_0=1, F_{-1}=P_0=Q_0=0$, where $a_{m,l}$,
$b_{m,l}$ and $c_{m,l}$($m, l\in {[0, 2,4, \cdots, n(n+1)]}$) are
unknown constants, $\mu$ and $\nu$ are the wave center.

Step4. Substituting Eq. (4) and Eq. (5) into Eq. (3) and equating
all the coefficients of the different powers of $\upsilon$ and $y$
to zero, we can know $a_{m,l}$, $b_{m,l}$ and $c_{m,l}$($m, l\in
{[0, 2,4, \cdots, n(n+1)]}$). The corresponding multiple rogue wave
solutions can be presented.

\section{1-rogue wave solutions} \label{sec:2}
\quad Based on the vcsca, set {\begin{eqnarray}\alpha(t)=\frac{6
\beta (t)}{\Theta _0}, \upsilon=x-\omega(t), u=2 \Theta
_0\,[ln\xi(\upsilon,y)]_{\upsilon\upsilon},\end{eqnarray} }Eq. (1)
can be changed as{\begin{eqnarray} &&6 \xi_\upsilon^2 [\xi
 [3 \beta (t)
\xi_{\upsilon\upsilon\upsilon\upsilon}-2 \omega '(t)
\xi_{\upsilon\upsilon}]+3 \beta
   (t) \xi_{\upsilon\upsilon}^2]+2 \xi^2 \xi_\upsilon [2 \omega '(t) \xi_{\upsilon\upsilon\upsilon}\nonumber\\&-&3 \beta (t)
   \xi_{\upsilon\upsilon\upsilon\upsilon\upsilon}]+\xi  [\xi
   [-3 \beta (t) \xi_{\upsilon\upsilon\upsilon\upsilon} \xi_{\upsilon\upsilon}+2 \beta
   (t) \xi_{\upsilon\upsilon\upsilon}^2+3 \omega '(t) \xi_{\upsilon\upsilon}^2]\nonumber\\&+&\xi^2
   [\beta (t) \xi_{\upsilon\upsilon\upsilon\upsilon\upsilon\upsilon}-\omega '(t)
   \xi_{\upsilon\upsilon\upsilon\upsilon}]-6 \beta (t) \xi_{\upsilon\upsilon}^3]-24 \beta (t) \xi_{\upsilon\upsilon\upsilon} \xi_\upsilon^3\nonumber\\&+&\gamma (t)
   [[6 \xi_\upsilon^2-2
   \xi  \xi_{\upsilon\upsilon}] \xi_y^2+2 \xi [\xi \xi_{\upsilon\upsilon y}-4
   \xi_\upsilon \xi_{\upsilon y}] \xi_y+\xi  [\xi_{yy}
   [\xi  \xi_{\upsilon\upsilon}-2 \xi_\upsilon^2]\nonumber\\&+&\xi  [2 \xi_{\upsilon y}^2+2 \xi_\upsilon \xi_{\upsilon yy}-\xi  \xi_{\upsilon\upsilon yy}]]]
   +6 \omega '(t) \xi_\upsilon^4.
\end{eqnarray}
}According to Eq. (5), we have {\begin{eqnarray}
\xi(\upsilon,y)=(\upsilon -\mu )^2+\zeta _1 (y-\nu )^2+\zeta _0,
\end{eqnarray}
}where $\mu$, $\nu$, $\zeta _0$ and $\zeta _1$ are unknown real
constants. Substituting Eq. (8) into Eq. (7) and equating the
coefficients of all powers $\upsilon$ and $y$ to zero, we obtain
\begin{eqnarray}
\gamma(t)=\frac{3 \beta (t)}{\zeta _0 \zeta _1}, \omega '(t)=\zeta
_1 \gamma (t).
\end{eqnarray}Substituting Eq. (8) and Eq. (9) into Eq. (6), the 1-rogue wave
solutions for Eq. (1) can be read as {\begin{eqnarray} u=\frac{4
\Theta _0 [-(\mu -\upsilon )^2+\zeta _1 (y-\nu )^2+\zeta
   _0]}{[(\mu -\upsilon )^2+\zeta_1 (y-\nu )^2+\zeta
   _0]{}^2}.
\end{eqnarray}}When $\zeta_0>0$, rogue wave
(10) has three extreme value points $(\mu, \nu)$, $(\mu \pm\sqrt{3}
\sqrt{\zeta _0}, \nu)$. When $\zeta_0< 0, \zeta_1> 0$, rogue wave
(10) has three extreme value points $(\mu, \nu)$, $(\mu, \nu
\pm\frac{\sqrt{-\zeta _0}}{\sqrt{\zeta _1}})$. Fig. 1 and Fig. 2
describe the dynamics features of rogue wave (10) when $\zeta_0$ and
$\zeta_1$ select different values.

\includegraphics[scale=0.55,bb=-20 270 10 10]{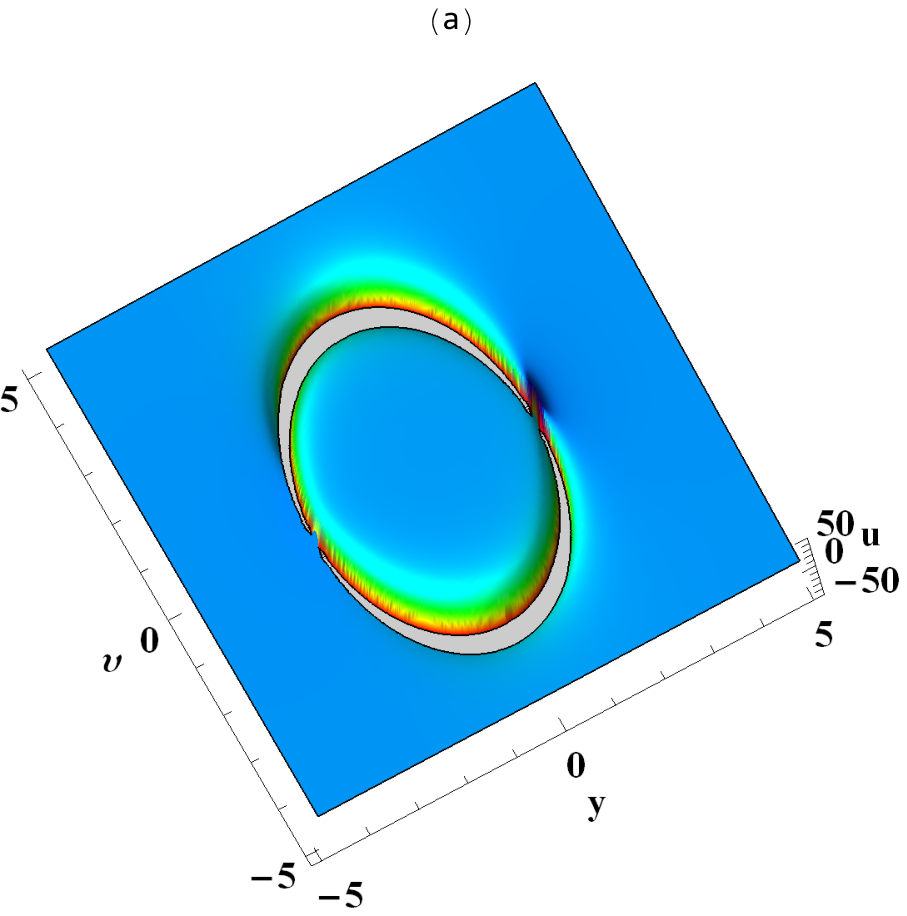}
\includegraphics[scale=0.45,bb=-270 310 10 10]{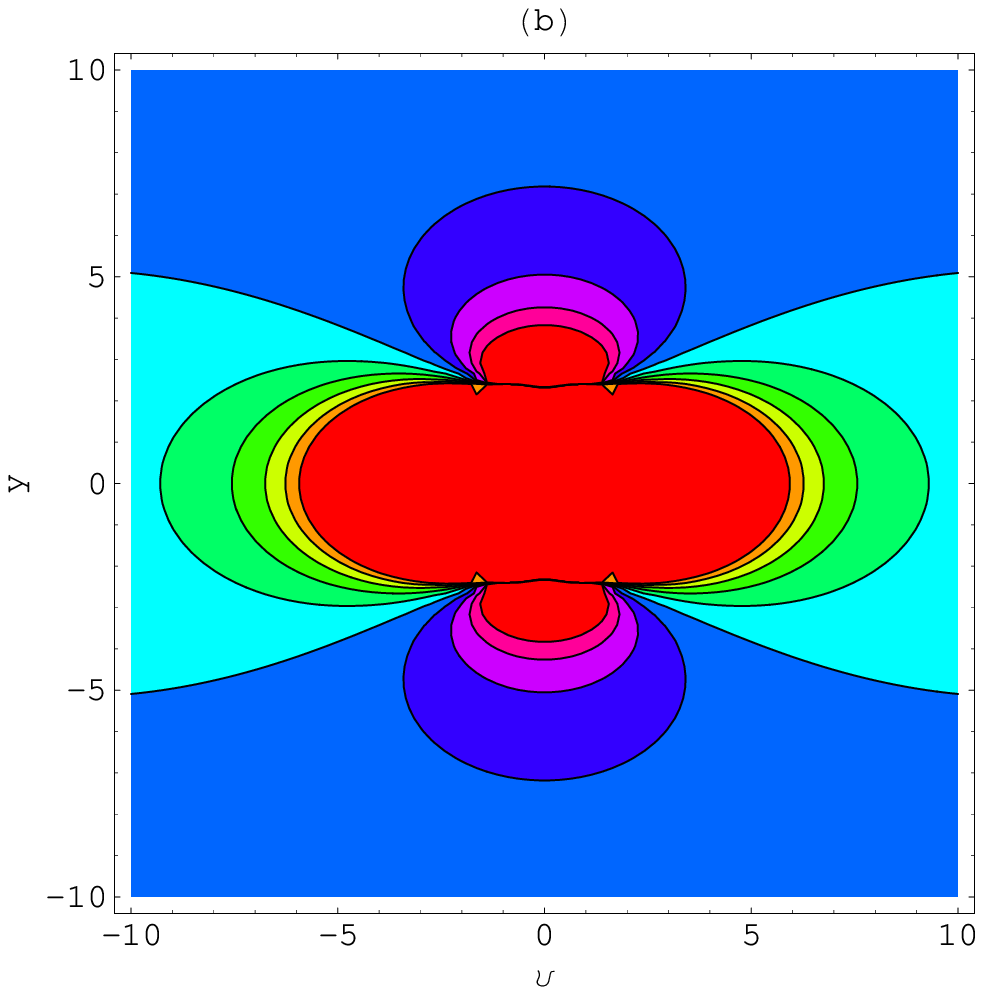}
\vspace{5.5cm}
\begin{tabbing}
\textbf{Fig. 1}. Rogue wave (10) with $\mu=\nu=0$, $\Theta _0=1$,
$\zeta_0=-10$, $\zeta_1=2$,\\ (a) 3D graphic,  (b) contour plot.
\end{tabbing}

\includegraphics[scale=0.55,bb=-20 270 10 10]{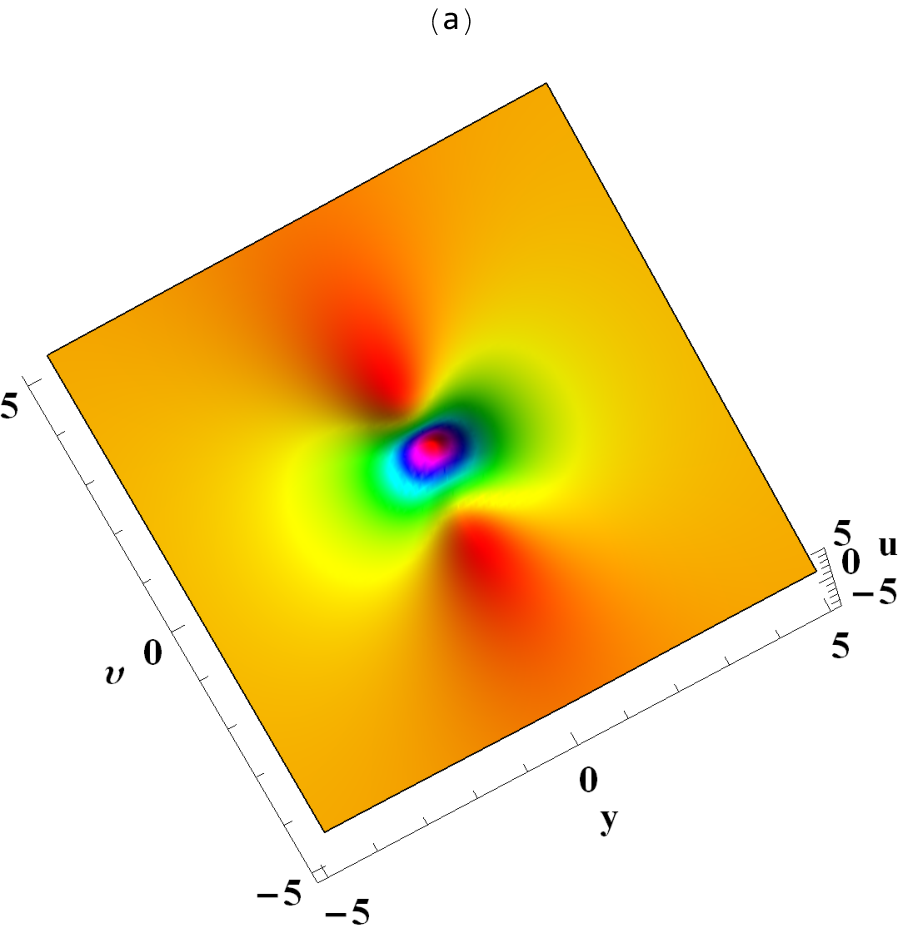}
\includegraphics[scale=0.45,bb=-270 310 10 10]{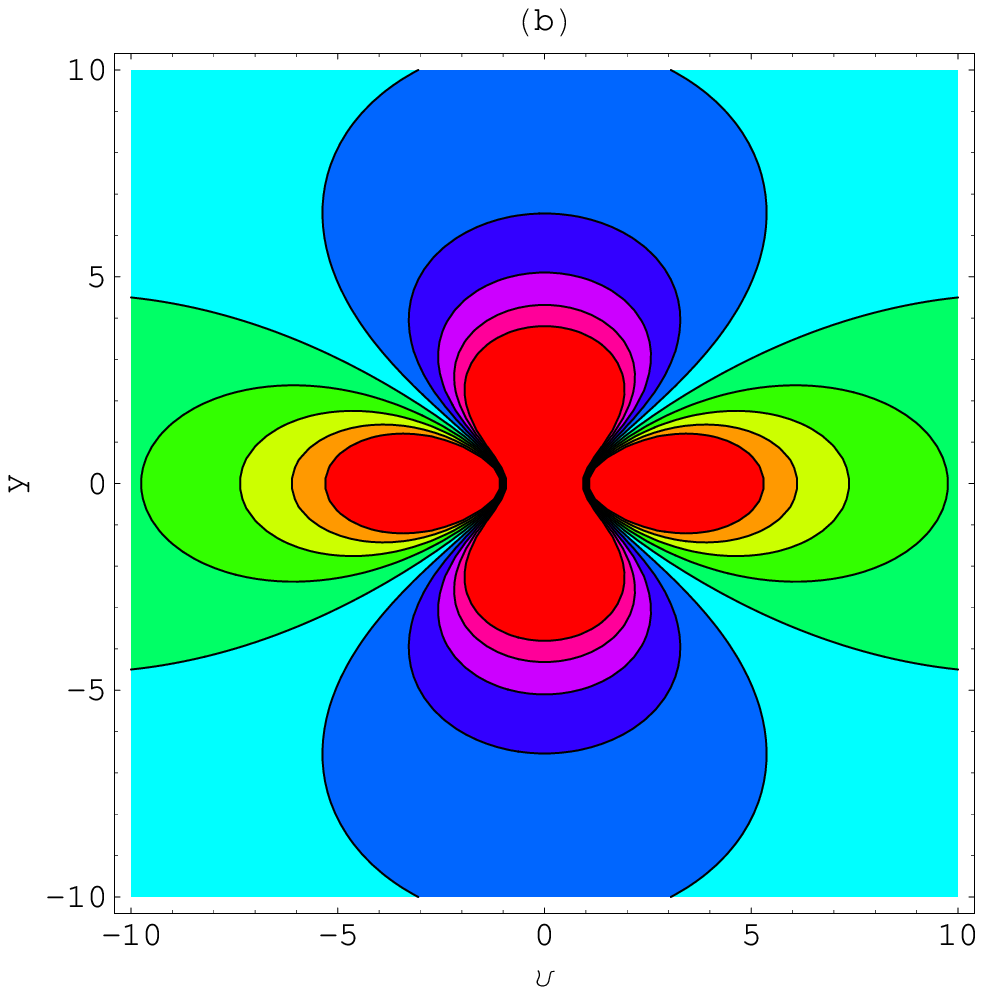}
\vspace{5.5cm}
\begin{tabbing}
\textbf{Fig. 2}. Rogue wave (10) with  $\mu=\nu=0$, $\Theta _0=1$,
$\zeta_0=1$, $\zeta_1=2$,\\ (a) 3D graphic, (b) contour plot.
\end{tabbing}

\section{3-rogue wave solutions} \label{sec:2}
In order to look for the 3-rogue wave solutions, we set
{\begin{eqnarray} \xi(\upsilon,y)&=&\mu ^2+\nu ^2+\upsilon ^6+y^6
\zeta _{17}+y^4 \zeta _{16}+2 \mu  \upsilon  \left(y^2
   \zeta _{23}+\upsilon ^2 \zeta _{24}+\zeta _{22}\right)\nonumber\\&+&2 \nu  y \left(y^2 \zeta
   _{20}+\upsilon ^2 \zeta _{21}+\zeta _{19}\right)+\upsilon ^4 y^2 \zeta _{11}+y^2
   \zeta _{15}\nonumber\\&+&\upsilon ^2 \left(y^4 \zeta _{14}+y^2 \zeta _{13}+\zeta
   _{12}\right)+\upsilon ^4 \zeta _{10}+\zeta _{18},
\end{eqnarray}
}where $\zeta _i (i=10,\cdots, 24)$  is unknown real constant.
Substituting Eq. (11) into Eq. (7) and equating the coefficients of
all powers $\upsilon$ and $y$ to zero, we get
\begin{eqnarray}\gamma(t)&=&\frac{90 \beta (t)}{\zeta _{13}}, \omega '(t)=\frac{30 \zeta _{11} \beta (t)}{\zeta _{13}},
\zeta_{14}=\frac{\zeta _{11}^2}{3}, \zeta_{16}=\frac{17 \zeta _{11} \zeta _{13}}{270},\nonumber\\
\zeta_{20}&=&-\frac{1}{9} \zeta _{11} \zeta _{21},
\zeta_{17}=\frac{\zeta _{11}^3}{27}, \zeta_{15}=\frac{19 \zeta
_{13}^2}{108 \zeta _{11}}, \zeta_{23}=-\zeta _{11}
\zeta _{24},\nonumber\\
\zeta_{22}&=&-\frac{\zeta _{13} \zeta _{24}}{30 \zeta _{11}},
\zeta_{12}=-\frac{5 \zeta _{13}^2}{36 \zeta _{11}^2},
\zeta_{10}=\frac{5 \zeta _{13}}{6 \zeta _{11}}, \zeta_{19}=\frac{\zeta _{13} \zeta _{21}}{18 \zeta _{11}},\nonumber\\
\zeta_{18}&=&-\mu ^2-\nu ^2+\mu ^2 \zeta _{24}^2+\frac{\nu ^2 \zeta
_{21}^2}{3
   \zeta _{11}}+\frac{5 \zeta _{13}^3}{72 \zeta _{11}^3}.
\end{eqnarray}Substituting Eq. (11) and Eq. (12) into Eq. (6), the 3-rogue wave
solutions for Eq. (1) can be read as {\begin{eqnarray} u&=&[24
\Theta _0 \zeta _{11} [5 [12 y^4 \zeta _{11}^4+36
   \zeta _{11}^2 \left(15 \upsilon ^4+y^2 \zeta _{13}+2 \nu  y \zeta
   _{21}+6 \mu  \upsilon  \zeta _{24}\right)\nonumber\\&+&216 \upsilon ^2 y^2 \zeta
   _{11}^3+180 \upsilon ^2 \zeta _{13} \zeta _{11}-5 \zeta
   _{13}^2] [40 y^6 \zeta _{11}^6+360 \upsilon ^2 y^4 \zeta
   _{11}^5\nonumber\\&+&2 \zeta _{11}^2 [95 y^2 \zeta _{13}^2+6 \zeta _{13}
   \left(75 \upsilon ^4+10 \nu  y \zeta _{21}-6 \mu  \upsilon  \zeta
   _{24}\right)+180 \nu ^2 \zeta _{21}^2]\nonumber\\&+&4 y^2 \zeta _{11}^4
   [y \left(17 y \zeta _{13}-60 \nu  \zeta _{21}\right)+270 \upsilon
   \left(\upsilon ^3-2 \mu  \zeta _{24}\right)]+1080 \zeta _{11}^3
   [\upsilon ^2 y^2 \zeta _{13}\nonumber\\&+&2 \nu  \upsilon ^2 y \zeta
   _{21}+\left(\upsilon ^3+\mu  \zeta _{24}\right){}^2]-150 \upsilon ^2
   \zeta _{13}^2 \zeta _{11}+75 \zeta _{13}^3]-12 \zeta
   _{11} [60 \upsilon  y^4 \zeta _{11}^4\nonumber\\&+&180 \upsilon  \zeta _{11}^2
   \left(3 \upsilon ^4+y^2 \zeta _{13}+2 \nu  y \zeta _{21}+3 \mu
   \upsilon  \zeta _{24}\right)+180 y^2 \zeta _{11}^3 \left(2 \upsilon
   ^3-\mu  \zeta _{24}\right)\nonumber\\&+&6 \zeta _{13} \zeta _{11} \left(50
   \upsilon ^3-\mu  \zeta _{24}\right)-25 \upsilon  \zeta
   _{13}^2]{}^2]]/[[40 y^6 \zeta _{11}^6+360 \upsilon ^2 y^4
   \zeta _{11}^5\nonumber\\&+&2 \zeta _{11}^2 [95 y^2 \zeta _{13}^2+6
   \zeta _{13} \left(75 \upsilon ^4+10 \nu  y \zeta _{21}-6 \mu  \upsilon
   \zeta _{24}\right)+180 \nu ^2 \zeta _{21}^2]\nonumber\\&+&4 y^2 \zeta
   _{11}^4 [y \left(17 y \zeta _{13}-60 \nu  \zeta _{21}\right)+270
   \upsilon  \left(\upsilon ^3-2 \mu  \zeta _{24}\right)]+1080 \zeta
   _{11}^3 [\upsilon ^2 y^2 \zeta _{13}\nonumber\\&+&2 \nu  \upsilon ^2 y \zeta
   _{21}+\left(\upsilon ^3+\mu  \zeta _{24}\right){}^2]-150 \upsilon ^2
   \zeta _{13}^2 \zeta _{11}+75 \zeta _{13}^3]{}^2],
\end{eqnarray}}where $\zeta _{11}$, $\zeta _{13}$, $\zeta _{21}$ and $\zeta _{24}$ are unrestricted.
Dynamics features of 3-rogue wave solutions are displayed in Figs.
3-6 when $(\mu,\nu)$ selects different values, we can see that three
rogue waves break apart and form a set of three 1-rogue waves in
Figs. 3-6.

\includegraphics[scale=0.55,bb=-20 270 10 10]{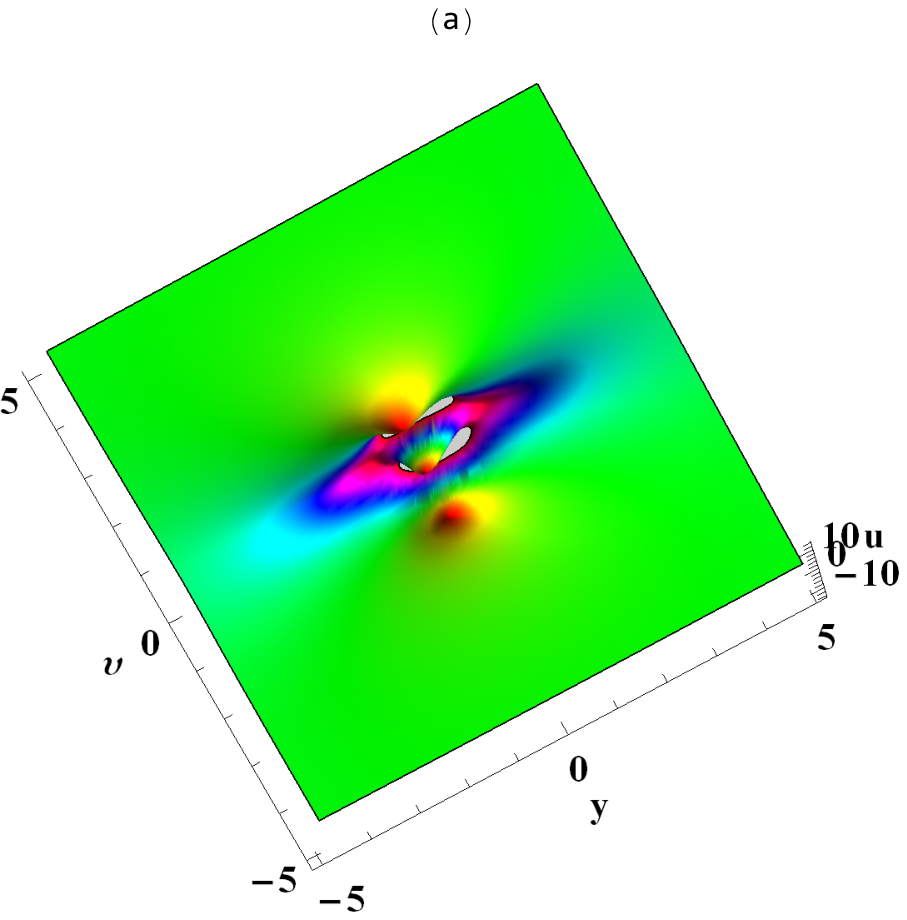}
\includegraphics[scale=0.45,bb=-270 310 10 10]{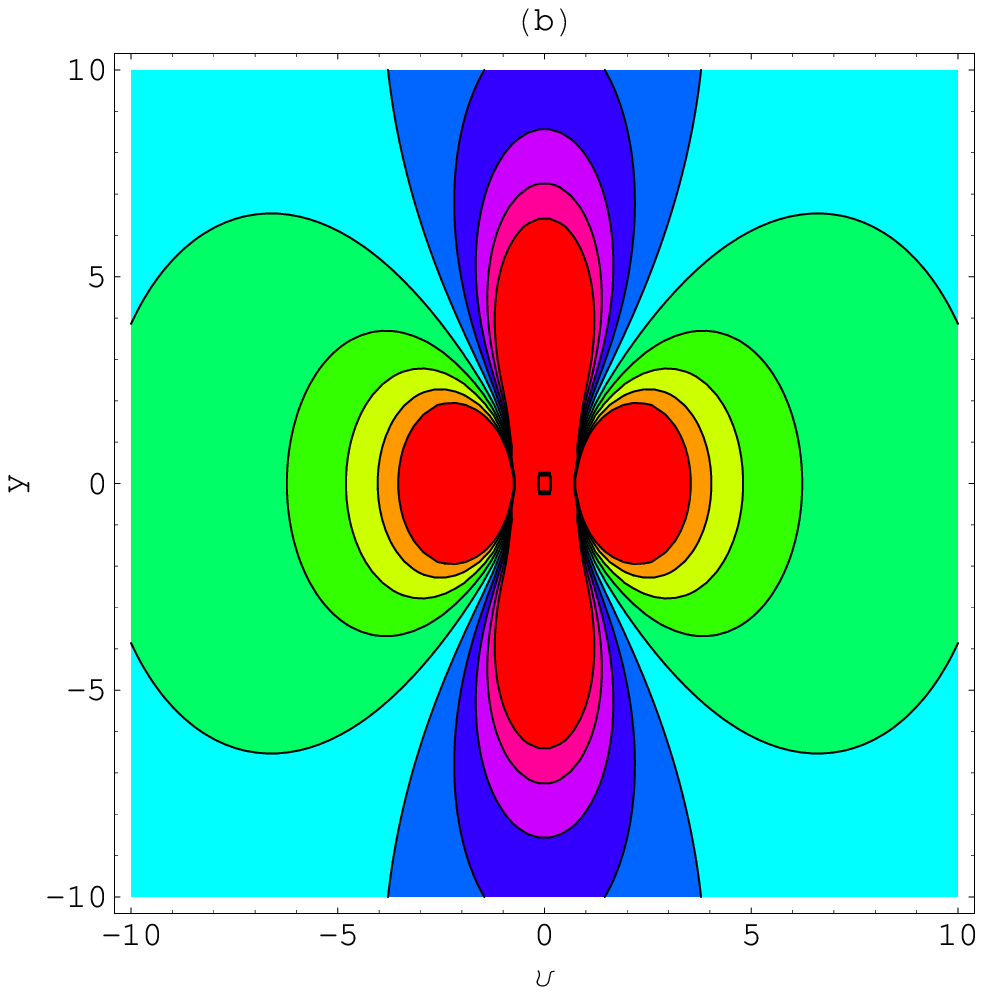}
\vspace{5.5cm}
\begin{tabbing}
\textbf{Fig. 3}. Rogue wave (13) with  $\mu=\nu=0$, $\Theta _0=1$,
$\zeta_{11}=\zeta_{13}=\zeta_{21}=\zeta_{24}=1$,\\ (a) 3D graphic,
(b) contour plot.
\end{tabbing}

\includegraphics[scale=0.55,bb=-20 270 10 10]{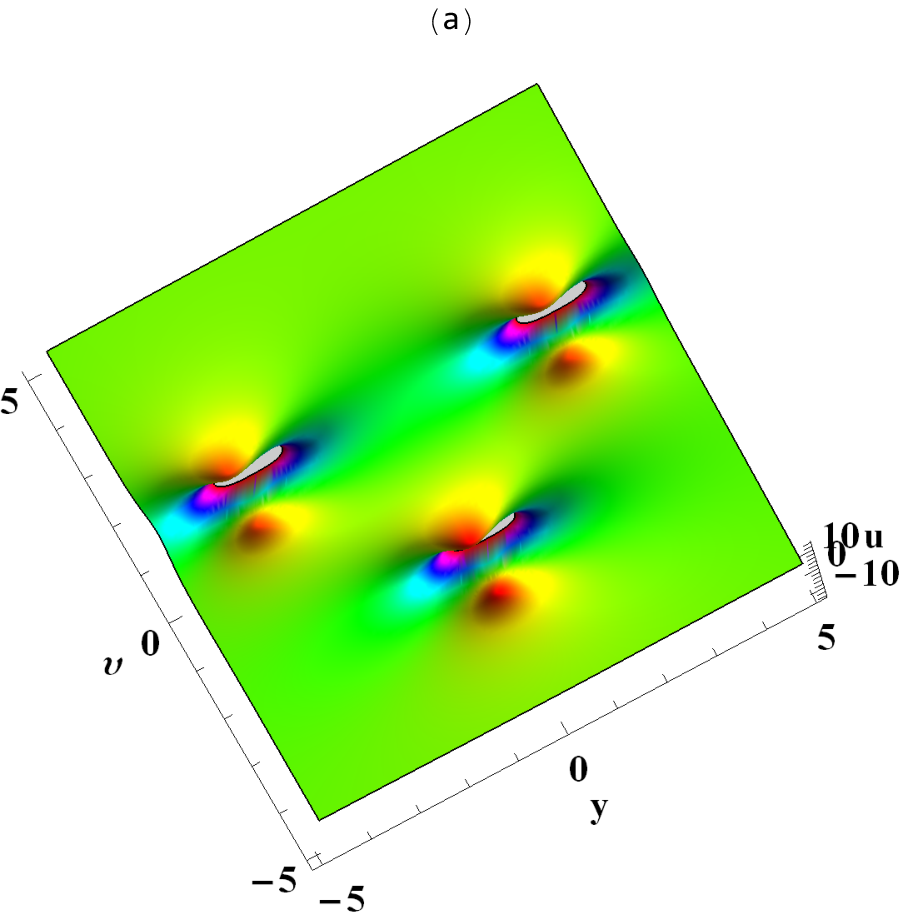}
\includegraphics[scale=0.45,bb=-270 310 10 10]{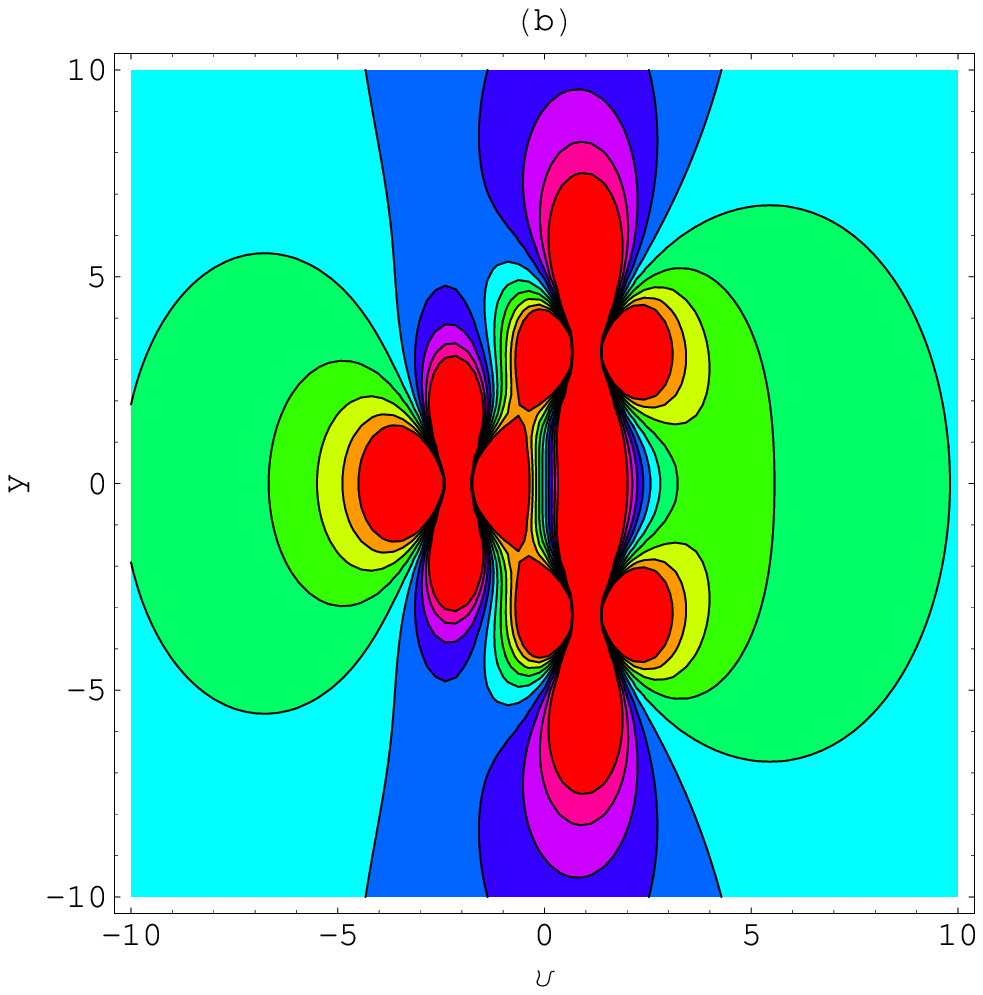}
\vspace{5.5cm}
\begin{tabbing}
\textbf{Fig. 4}. Rogue wave (13) with  $\mu=10, \nu=0$,
$\zeta_{11}=\zeta_{13}=\zeta_{21}=\zeta_{24}=1$,\\ $\Theta _0=1$,
(a) 3D graphic, (b) contour plot.
\end{tabbing}

\includegraphics[scale=0.55,bb=-20 270 10 10]{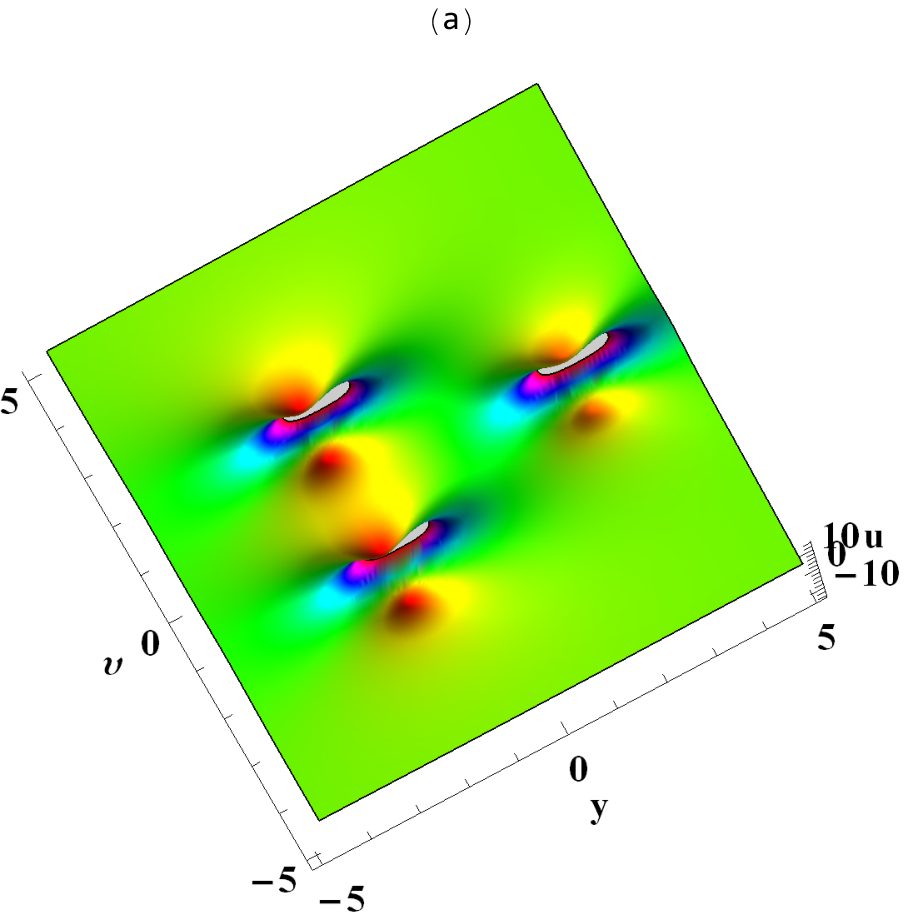}
\includegraphics[scale=0.45,bb=-270 310 10 10]{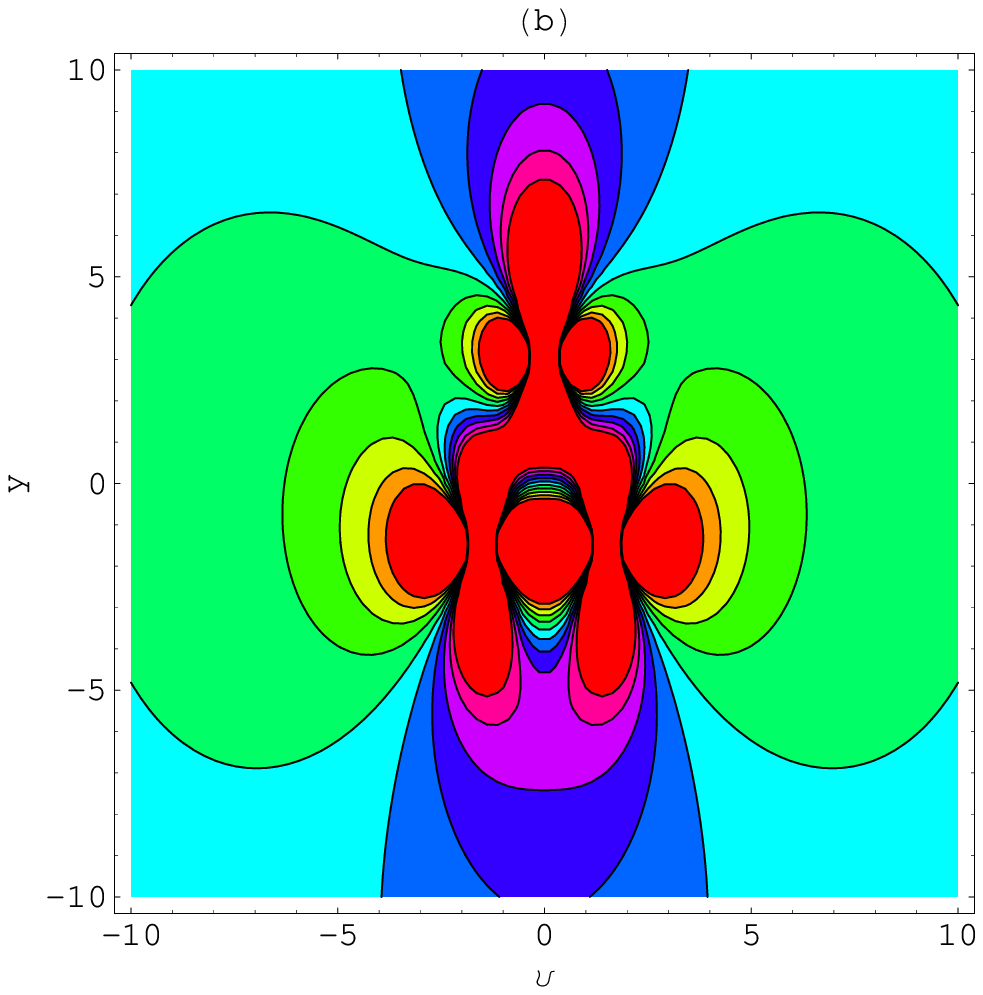}
\vspace{5.5cm}
\begin{tabbing}
\textbf{Fig. 5}. Rogue wave (13) with  $\mu=0, \nu=10$,
$\zeta_{11}=\zeta_{13}=\zeta_{21}=\zeta_{24}=1$,\\ $\Theta _0=1$,
(a) 3D graphic, (b) contour plot.
\end{tabbing}

\includegraphics[scale=0.55,bb=-20 270 10 10]{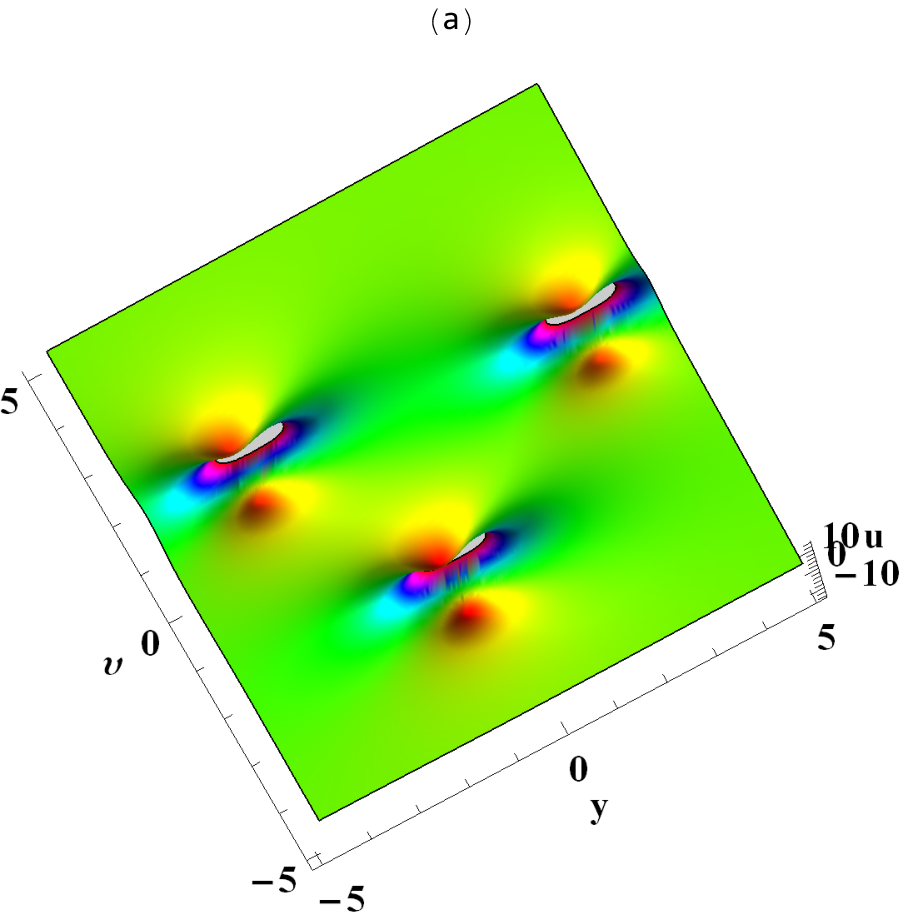}
\includegraphics[scale=0.45,bb=-270 310 10 10]{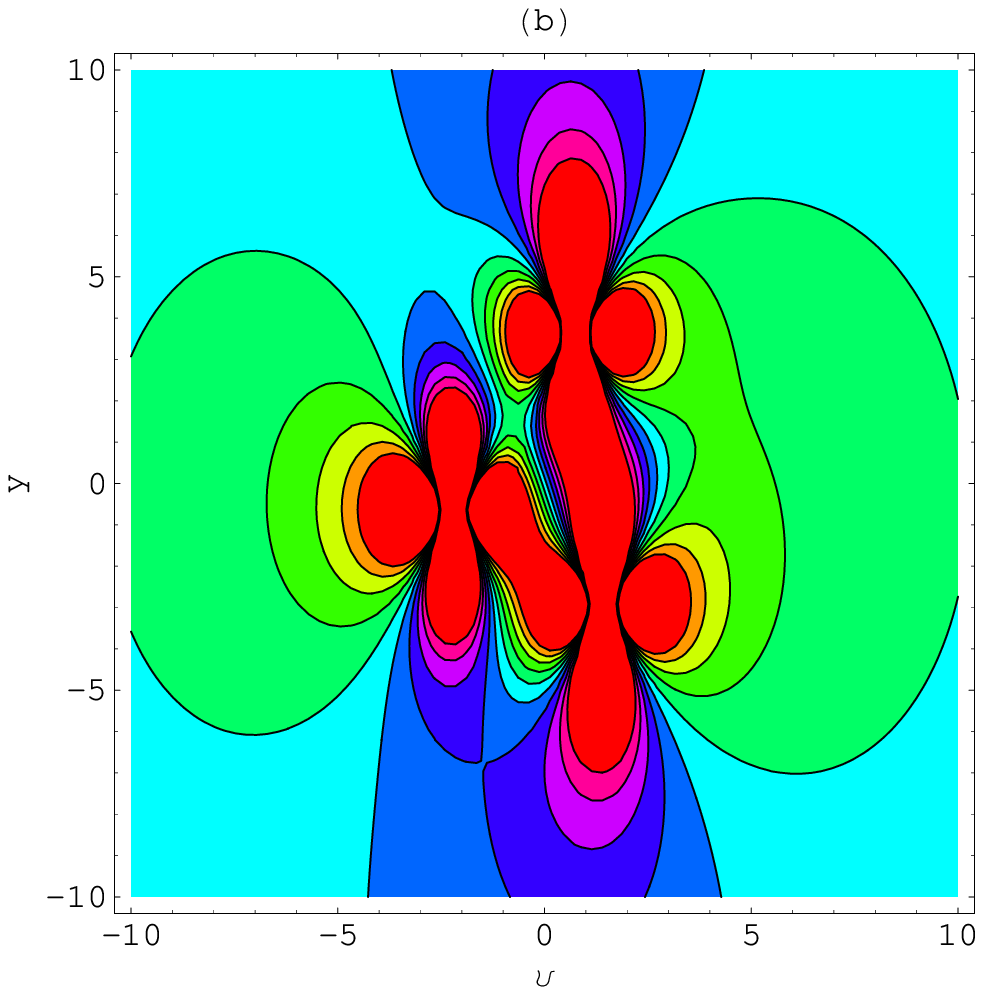}
\vspace{5.5cm}
\begin{tabbing}
\textbf{Fig. 6}. Rogue wave (13) with  $\mu=\nu=10$,
$\zeta_{11}=\zeta_{13}=\zeta_{21}=\zeta_{24}=1$,\\ $\Theta _0=1$,
(a) 3D graphic, (b) contour plot.
\end{tabbing}

\section{6-rogue wave solutions} \label{sec:2}
To present the 6-rogue wave solutions, we assume {\begin{eqnarray}
\xi(\upsilon,y)&=&\upsilon ^{12}+y^8 \zeta _{48}+y^6 \zeta
   _{47}+y^4 \zeta _{46}+\upsilon ^{10}
   \left(y^2 \zeta _{26}+\zeta _{25}\right)\nonumber\\&+&y^2 \zeta _{45}+\upsilon ^8 \left(y^4
   \zeta _{29}+y^2 \zeta _{28}+\zeta _{27}\right)+2 \mu  \upsilon  [\upsilon ^6+y^6
   \zeta _{64}+y^4 \zeta _{63}\nonumber\\&+&\upsilon ^4 \left(y^2 \zeta _{69}+\zeta
   _{68}\right)+y^2 \zeta _{62}+\upsilon ^2 \left(y^4 \zeta _{67}+y^2 \zeta
   _{66}+\zeta _{65}\right)+\zeta _{61}]\nonumber\\&+&2 \nu  y [y^6+y^4 \left(\upsilon ^2
   \zeta _{57}+\zeta _{56}\right)+y^2 \left(\upsilon ^4 \zeta _{55}+\upsilon ^2
   \zeta _{54}+\zeta _{53}\right)+\upsilon ^6 \zeta _{60}\nonumber\\&+&\upsilon ^4 \zeta
   _{59}+\upsilon ^2 \zeta _{58}+\zeta _{52}]+\upsilon ^6 \left(y^6 \zeta
   _{33}+y^4 \zeta _{32}+y^2 \zeta _{31}+\zeta _{30}\right)\nonumber\\&+&\upsilon ^4 \left(y^8
   \zeta _{38}+y^6 \zeta _{37}+y^4 \zeta _{36}+y^2 \zeta _{35}+\zeta
   _{34}\right)+\upsilon ^2 (y^{10} \zeta _{44}+y^8 \zeta _{43}\nonumber\\&+&y^6 \zeta _{42}+y^4
   \zeta _{41}+y^2 \zeta _{40}+\zeta _{39})+\zeta _{51}+y^{12} \zeta _{50}+y^{10}
\zeta _{49}\nonumber\\&+&\left(\mu ^2+\nu ^2\right) [\upsilon ^2+y^2
\zeta _1+\zeta _0],
\end{eqnarray}
}where $\zeta _i (i=25,\cdots, 69)$  is unknown real constant.
Substituting Eq. (14) into Eq. (7) and equating the coefficients of
all powers $\upsilon$ and $y$ to zero, we obtain
\begin{eqnarray}\gamma(t)&=&\frac{690 \beta (t)}{\zeta _{28}}, \omega '(t)=\frac{1}{6} \zeta _{26} \gamma (t),
\zeta_{29}=\frac{5 \zeta _{26}^2}{12}, \zeta_{33}=\frac{5 \zeta
_{26}^3}{54},\nonumber\\ \zeta_{32}&=&\frac{77 \zeta _{26} \zeta
_{28}}{207}, \zeta_{31}=\frac{1862 \zeta _{28}^2}{7935 \zeta _{26}},
\zeta_{37}=\frac{73 \zeta _{26}^2 \zeta _{28}}{1242},
\zeta_{36}=\frac{749 \zeta _{28}^2}{9522},\nonumber\\
\zeta_{55}&=&-\frac{180}{\zeta _{26}^2}, \zeta_{38}=\frac{5 \zeta
_{26}^4}{432},
\zeta_{35}=\frac{294 \zeta _{28}^3}{12167 \zeta _{26}^2}, \zeta_{43}=\frac{19 \zeta _{26}^3 \zeta _{28}}{4968},\nonumber\\
\zeta_{42}&=&\frac{77 \zeta _{26} \zeta _{28}^2}{6210},
\zeta_{41}=-\frac{49 \zeta _{28}^3}{182505 \zeta _{26}}, \zeta_{52}=\frac{271656 \zeta _{28}^3}{304175 \zeta _{26}^6},\nonumber\\
\zeta_{54}&=&-\frac{1368 \zeta _{28}}{23 \zeta _{26}^3},
\zeta_{44}=\frac{\zeta _{26}^5}{1296},
\zeta_{57}=-\frac{54}{\zeta _{26}}, \zeta_{40}=\frac{3773 \zeta _{28}^4}{6996025 \zeta _{26}^3},\nonumber\\
\zeta_{50}&=&\frac{\zeta _{26}^6}{46656}, \zeta_{49}=\frac{29 \zeta
_{26}^4 \zeta _{28}}{447120},
\zeta_{48}=\frac{289 \zeta _{26}^2 \zeta _{28}^2}{1142640},\nonumber\\
\zeta_{64}&=&\frac{5 \zeta _{26}^3}{216}, \zeta_{47}=\frac{39949
\zeta _{28}^3}{49276350},
\zeta_{27}=\frac{147 \zeta _{28}^2}{2645 \zeta _{26}^2},\nonumber\\
\zeta_{67}&=&-\frac{5 \zeta _{26}^2}{36},
\zeta_{63}=\frac{\zeta _{26} \zeta _{28}}{92}, \zeta_{66}=-\frac{\zeta _{28}}{3},\nonumber\\
\zeta_{25}&=&\frac{98 \zeta _{28}}{115 \zeta _{26}},
\zeta_{56}=-\frac{42 \zeta _{28}}{115 \zeta _{26}^2},
\zeta_{46}=\frac{655669 \zeta _{28}^4}{755570700 \zeta
_{26}^2},\nonumber\\ \zeta_{53}&=&-\frac{1764 \zeta _{28}^2}{2645
\zeta _{26}^4}, \zeta_{60}=\frac{1080}{\zeta _{26}^3},
\zeta_{69}=-\frac{3 \zeta _{26}}{2}, \zeta_{65}=-\frac{49 \zeta _{28}^2}{2645 \zeta _{26}^2},\nonumber\\
\zeta_{30}&=&\frac{15092 \zeta _{28}^3}{912525 \zeta _{26}^3},
\zeta_{39}=-\nu ^2+\frac{279936 \nu ^2}{\zeta _{26}^7}+\frac{6391462
\zeta
   _{28}^5}{2413628625 \zeta _{26}^5},\nonumber\\
\zeta_{34}&=&-\frac{41503 \zeta _{28}^4}{4197615 \zeta _{26}^4},
\zeta_{68}=\frac{13 \zeta _{28}}{115 \zeta _{26}},
\zeta_{58}=-\frac{28728 \zeta _{28}^2}{2645 \zeta _{26}^5},\nonumber\\
\zeta_{45}&=&-\zeta _1 \left(\mu ^2+\nu ^2\right)+\frac{\mu ^2 \zeta
   _{26}}{6}+\frac{46656 \nu ^2}{\zeta _{26}^6}+\frac{1203587 \zeta
   _{28}^5}{1448177175 \zeta _{26}^4},
\nonumber\\
\zeta_{62}&=&\frac{107 \zeta _{28}^2}{15870 \zeta _{26}},
\zeta_{61}=\frac{2401 \zeta _{28}^3}{912525 \zeta _{26}^3},  \zeta_{59}=\frac{4536 \zeta _{28}}{23 \zeta _{26}^4},\nonumber\\
\zeta_{51}&=&\frac{3 \zeta _{28} \left(279936 \nu ^2+\mu ^2 \zeta
_{26}^7\right)}{115
   \zeta _{26}^8}-\zeta _0 \left(\mu ^2+\nu ^2\right)+\frac{35153041
   \zeta _{28}^6}{832701875625 \zeta _{26}^6}.
\end{eqnarray}Substituting Eq. (14) and Eq. (15) into Eq. (6), the 6-rogue wave
solutions for Eq. (1) can be written as {\begin{eqnarray} u=2 \Theta
_0
\left(\frac{\xi_{\upsilon\upsilon}}{\xi}-\frac{\xi_\upsilon^2}{\xi^2}\right),
\end{eqnarray}}where $\xi$ satisfies Eq. (14) and Eq. (15), $\zeta _{26}$ and $\zeta _{28}$ are unrestricted.
Dynamics features of 6-rogue wave solutions are shown in Figs. 7-10
when $(\mu,\nu)$ selects different values, we can see that sixrogue
waves break apart and form a set of six 1-rogue waves in Figs. 7-10.

\includegraphics[scale=0.55,bb=-20 270 10 10]{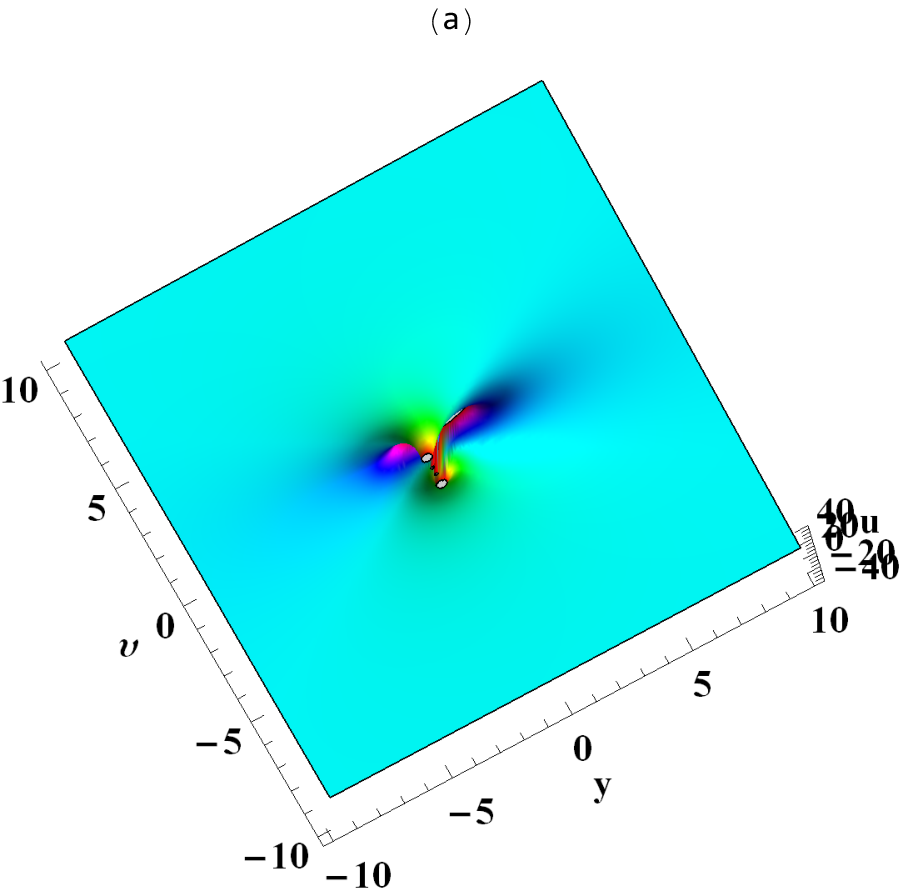}
\includegraphics[scale=0.45,bb=-270 310 10 10]{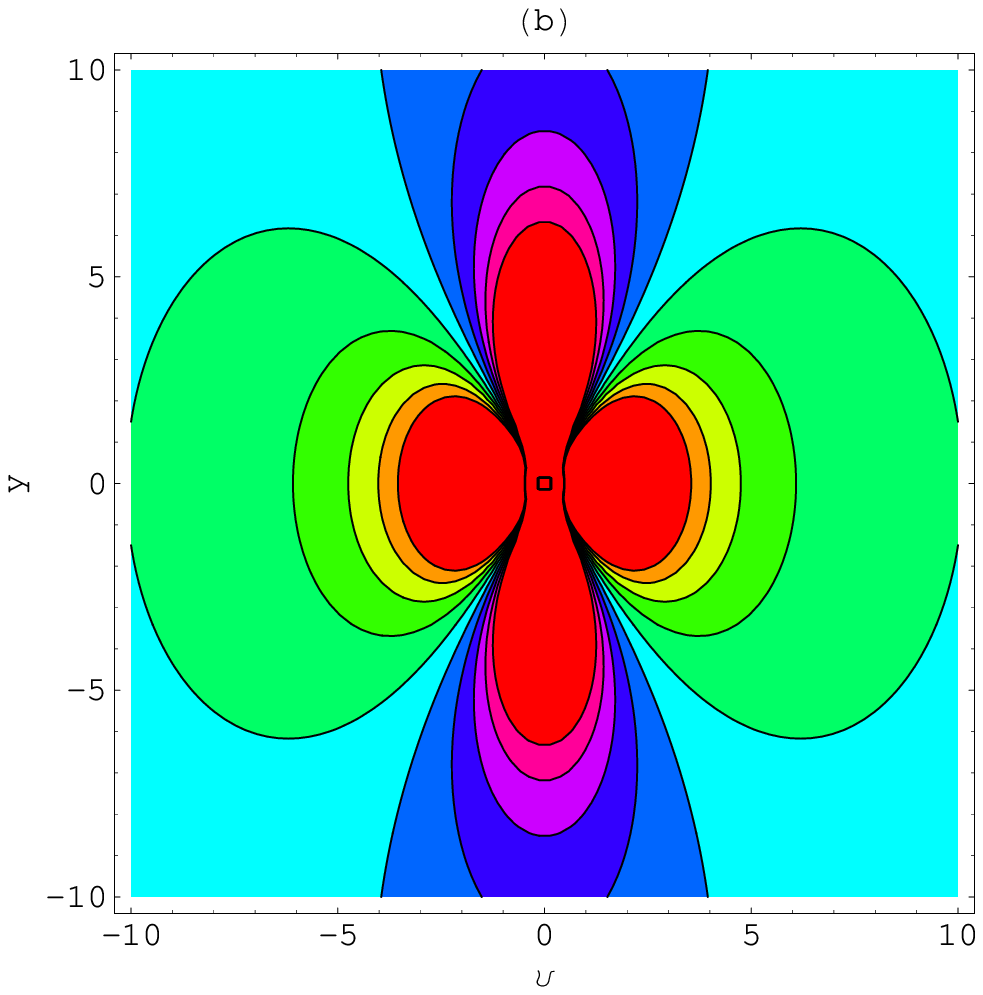}
\vspace{5.5cm}
\begin{tabbing}
\textbf{Fig. 7}. Rogue wave (16) with $\mu=\nu=0$,  $\Theta _0=1$,
$\zeta_0=\zeta_1=\zeta_{28}=1$,\\ $\zeta_{26}=2$, (a) 3D graphic,
(b) contour plot.
\end{tabbing}

\includegraphics[scale=0.55,bb=-20 270 10 10]{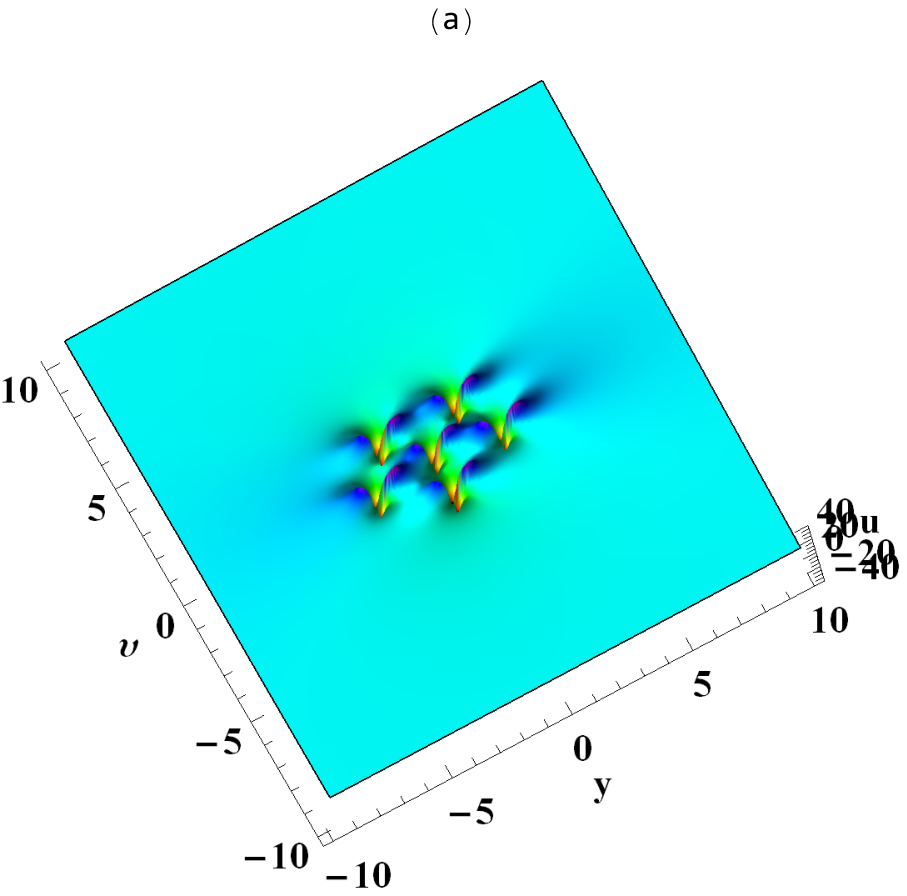}
\includegraphics[scale=0.45,bb=-270 310 10 10]{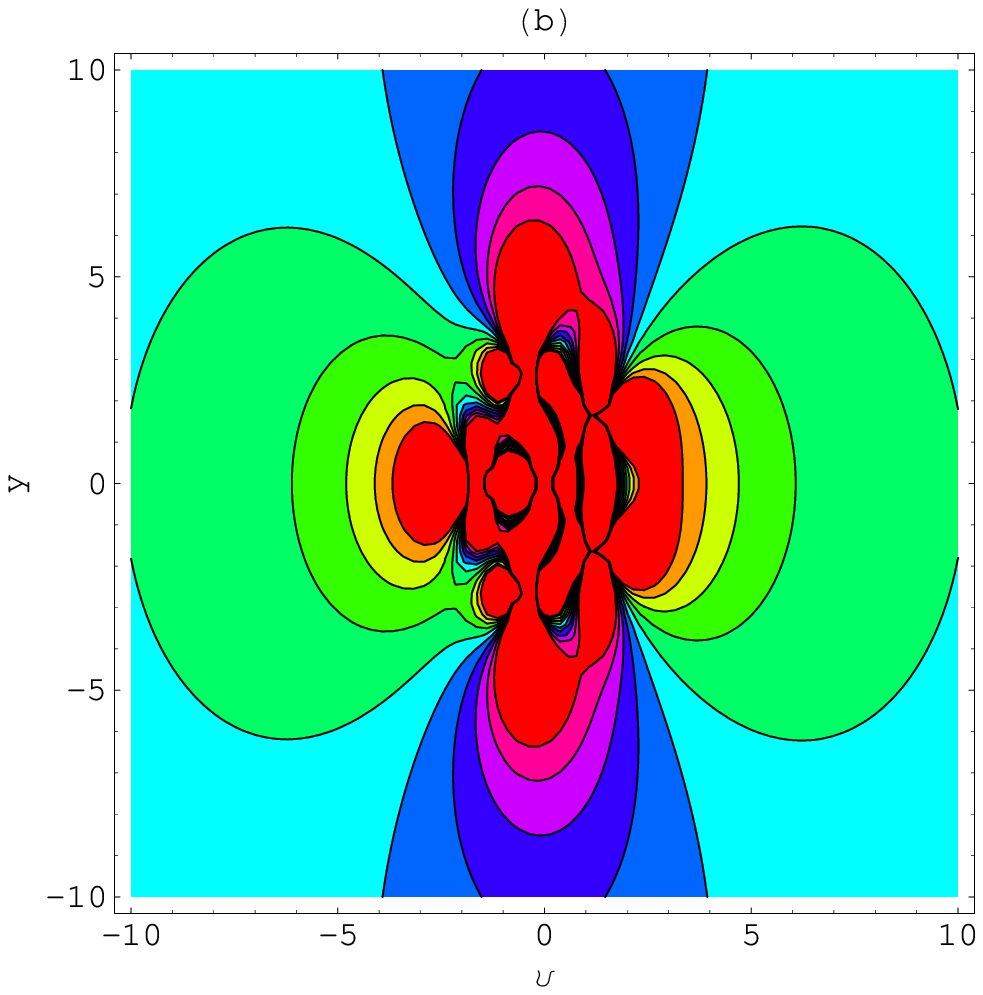}
\vspace{5.5cm}
\begin{tabbing}
\textbf{Fig. 8}. Rogue wave (16) with $\mu=10, \nu=0$, $\Theta
_0=1$, $\zeta_0=\zeta_1=\zeta_{28}=1$,\\ $\zeta_{26}=2$, (a) 3D
graphic,  (b) contour plot.
\end{tabbing}

\includegraphics[scale=0.55,bb=-20 270 10 10]{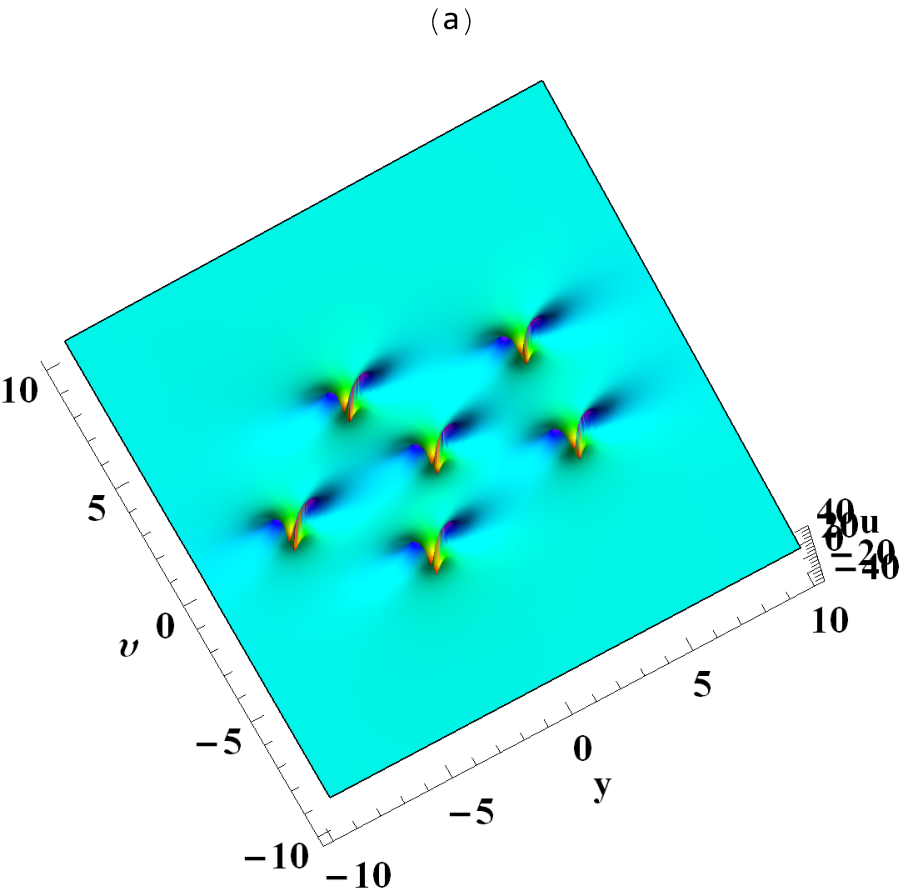}
\includegraphics[scale=0.45,bb=-270 310 10 10]{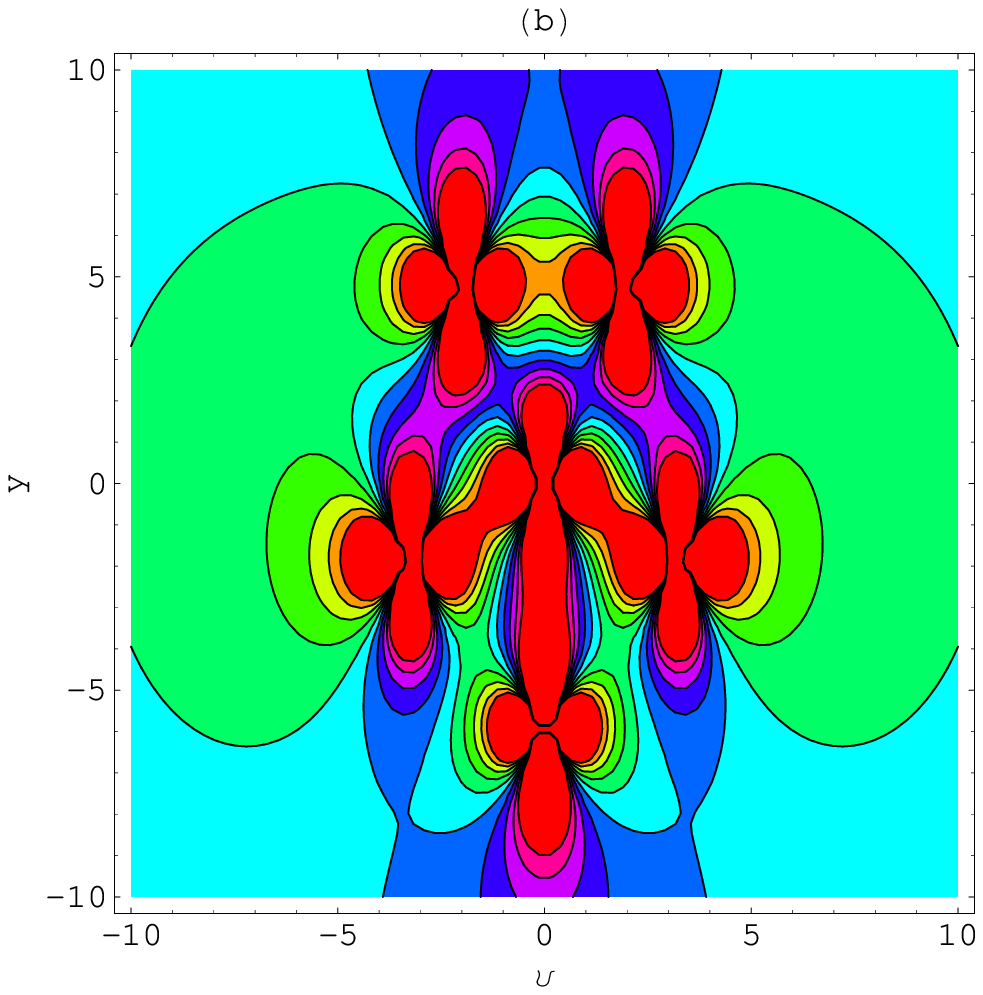}
\vspace{5.5cm}
\begin{tabbing}
\textbf{Fig. 9}. Rogue wave (16) with $\mu=0, \nu=10$, $\Theta
_0=1$, $\zeta_0=\zeta_1=\zeta_{28}=1$,\\ $\zeta_{26}=2$, (a) 3D
graphic,  (b) contour plot.
\end{tabbing}

\includegraphics[scale=0.55,bb=-20 270 10 10]{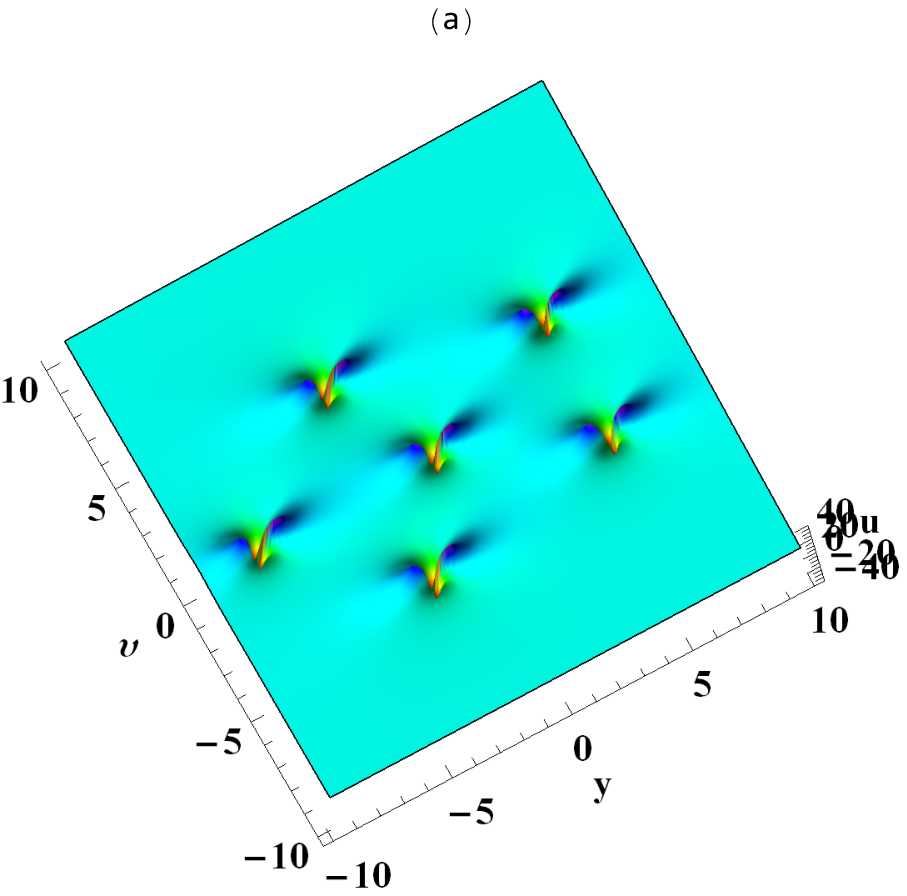}
\includegraphics[scale=0.45,bb=-270 310 10 10]{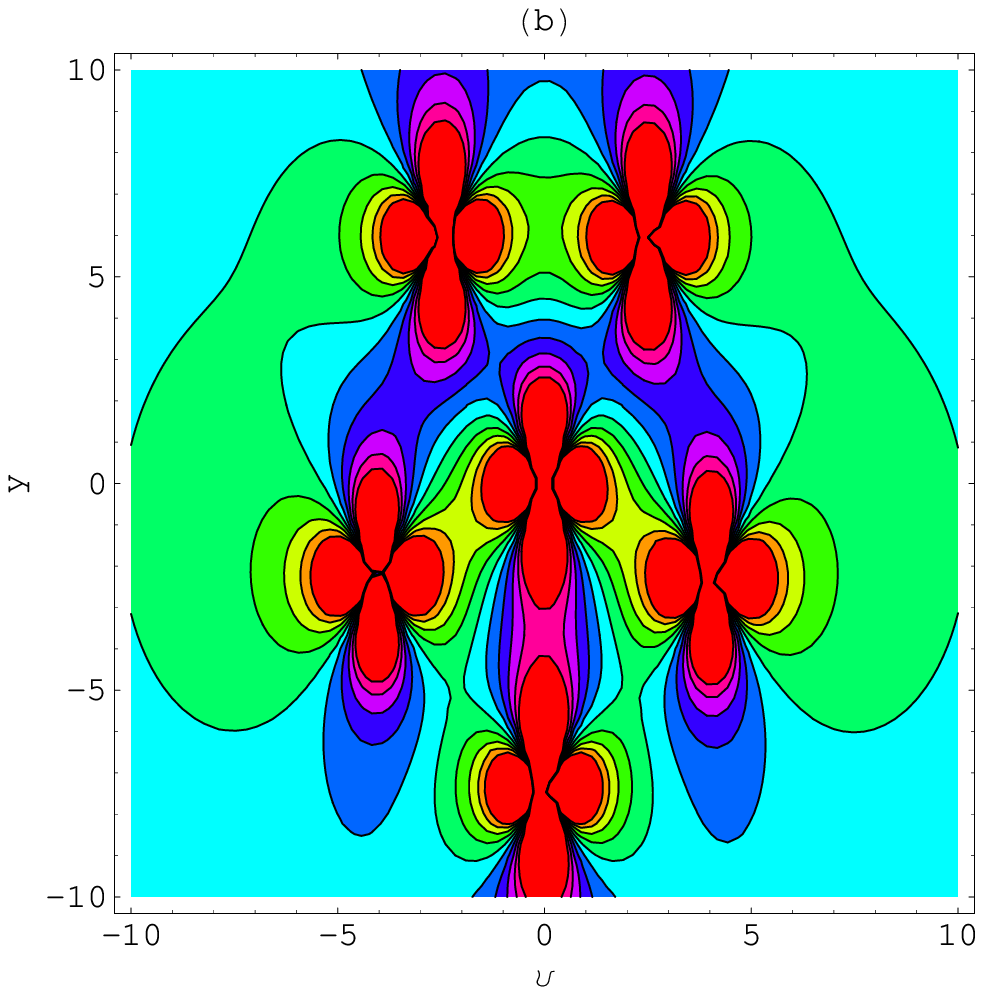}
\vspace{5.5cm}
\begin{tabbing}
\textbf{Fig. 10}. Rogue wave (16) with $\mu=\nu=30$, $\Theta _0=1$,
$\zeta_0=\zeta_1=\zeta_{28}=1$,\\ $\zeta_{26}=2$, (a) 3D graphic,
(b) contour plot.
\end{tabbing}

\section{ Conclusion}
\label{sec:3} \quad In the paper,   a variable-coefficient symbolic
computation approach is proposed. The main difference between this
method and the previous one in Ref. [15] is that we replace the
traveling wave transformation with the non-traveling wave
transformation, making it suitable for solving the multiple rogue
wave solution of the nonlinear system with variable coefficients.
This change has not been seen in other literatures. Applied the
vcsca to the (2+1)-dimensional vcKP equation, the 1-rogue wave
solutions, 3-rogue wave solutions and 6-rogue wave solutions are
present. By setting different values of $(\mu, \nu)$,
 their dynamics features are displayed in Figs. 1-10. All the obtained solutions have been verified to be correct.

 \quad Substituting $\upsilon=x-\omega(t)$ in rogue wave solution (10), we have\\
{\begin{eqnarray} u(x,y,z,t)=\frac{4 \Theta _0 [-[\mu +\frac{3 \int
\beta (t) \, dt}{\zeta
   _0}-x]{}^2+\zeta _1 (y-\nu )^2+\zeta _0]}{[[\mu
   +\frac{3 \int \beta (t) \, dt}{\zeta _0}-x]{}^2+\zeta _1 (y-\nu
   )^2+\zeta _0]{}^2}.
\end{eqnarray}}When variable-coefficient $\beta (t)$ chooses
different function, the rogue wave (17) shows different dynamics
features in Fig. 11.

\includegraphics[scale=0.4,bb=20 270 10 10]{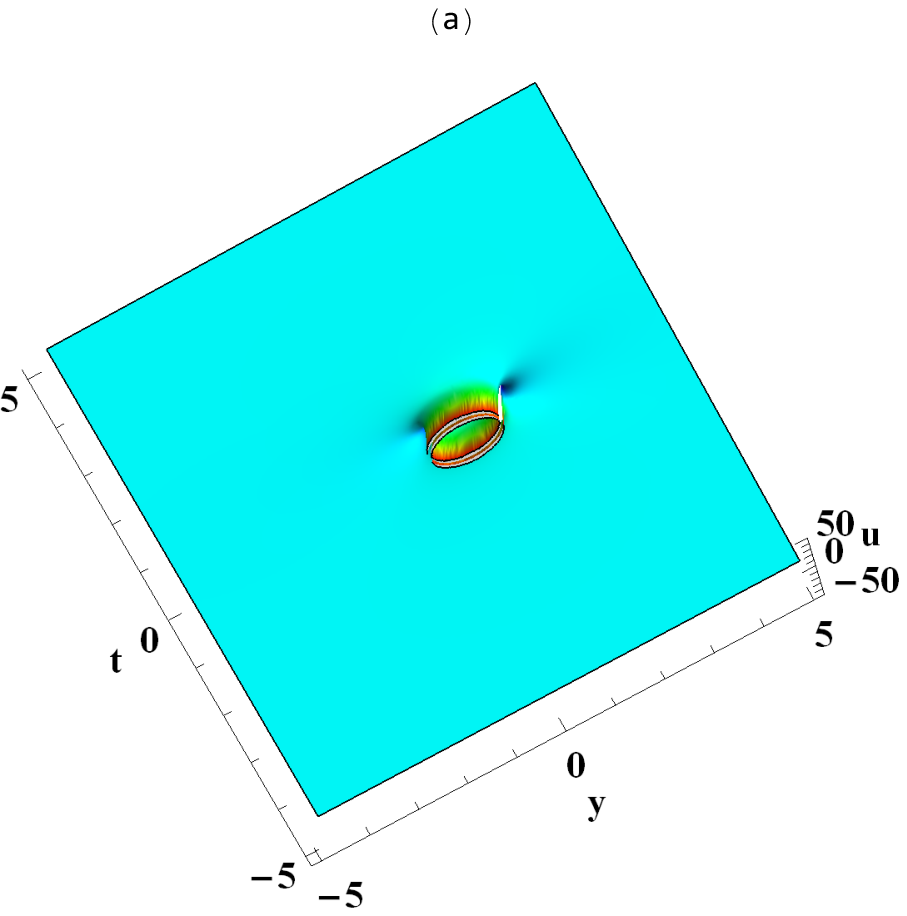}
\includegraphics[scale=0.4,bb=-255 270 10 10]{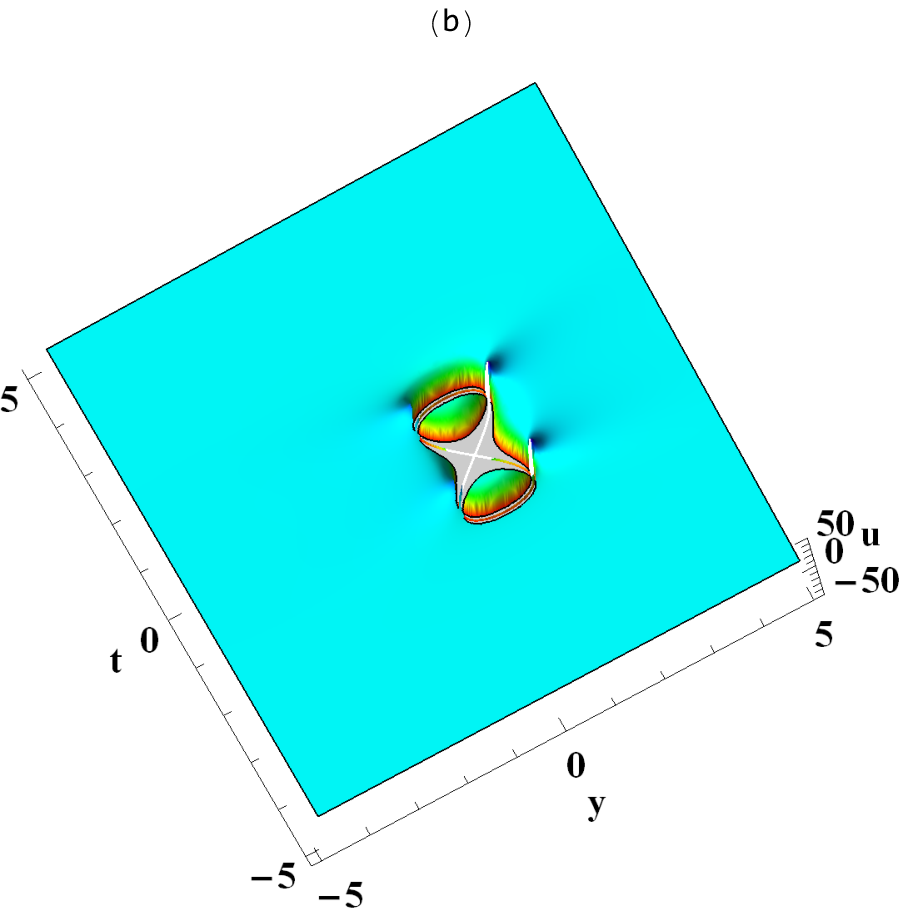}
\includegraphics[scale=0.4,bb=-260 270 10 10]{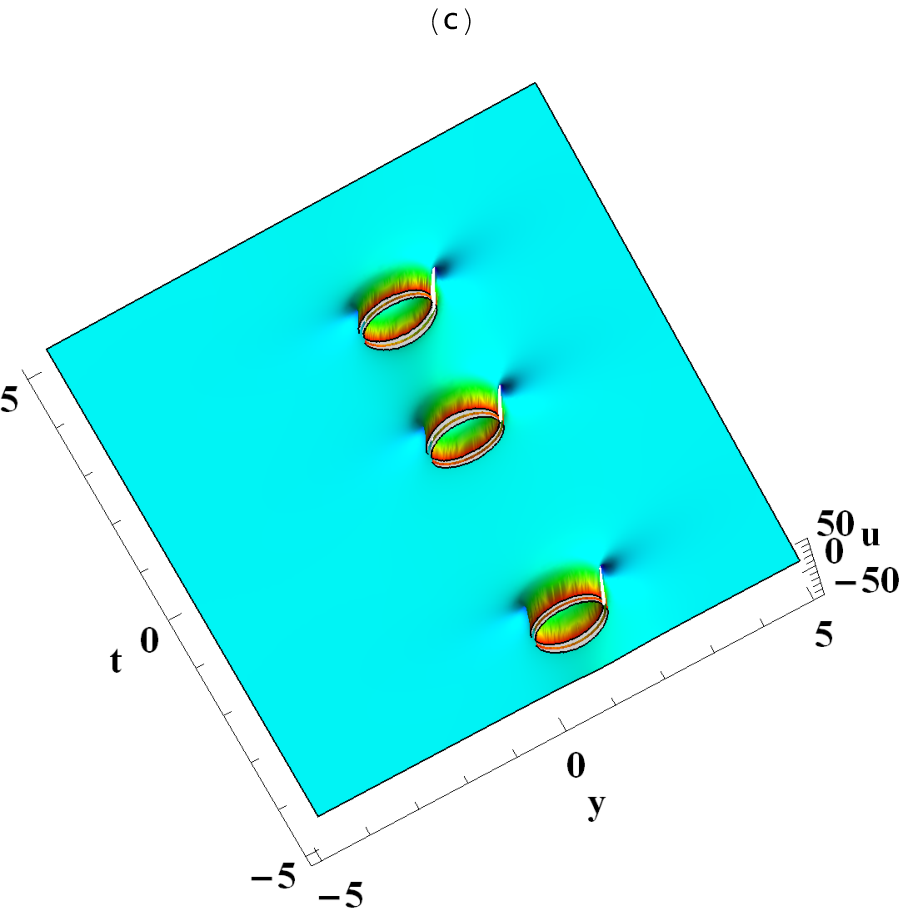}
\includegraphics[scale=0.35,bb=750 620 10 10]{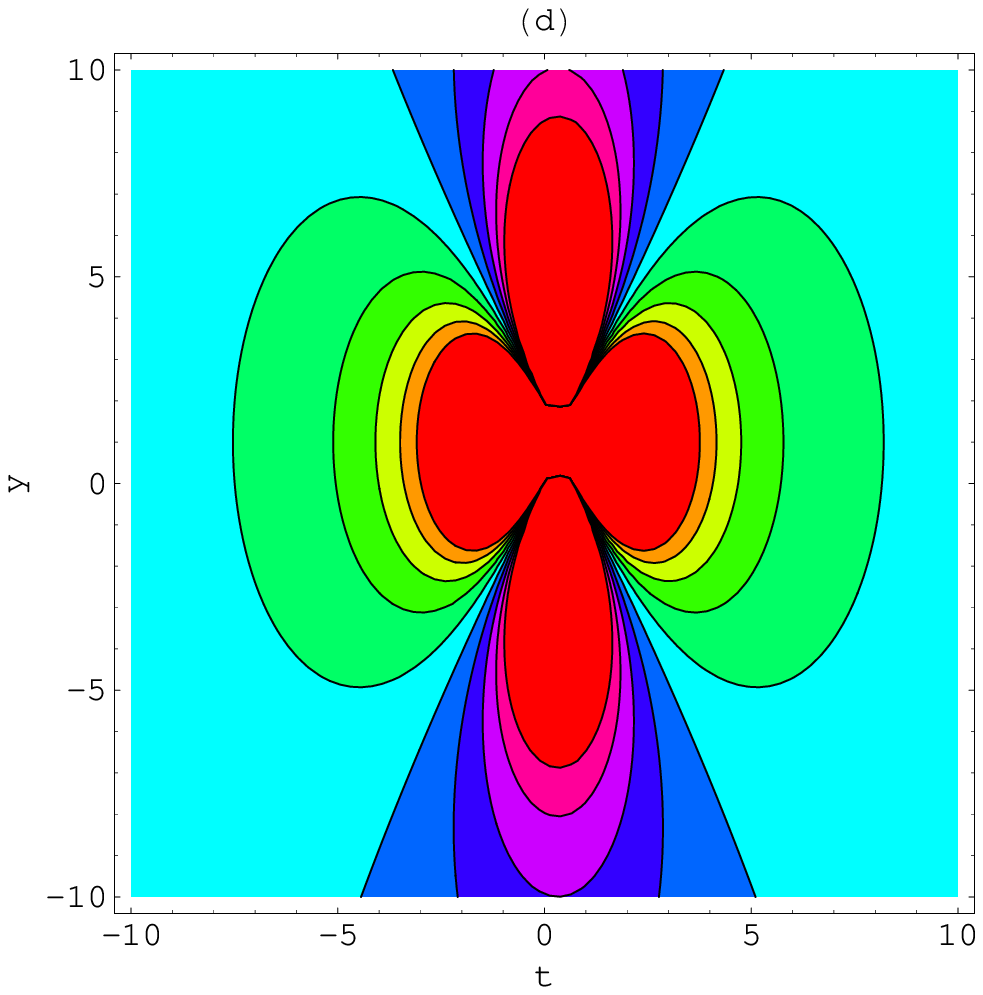}
\includegraphics[scale=0.35,bb=-290 620 10 10]{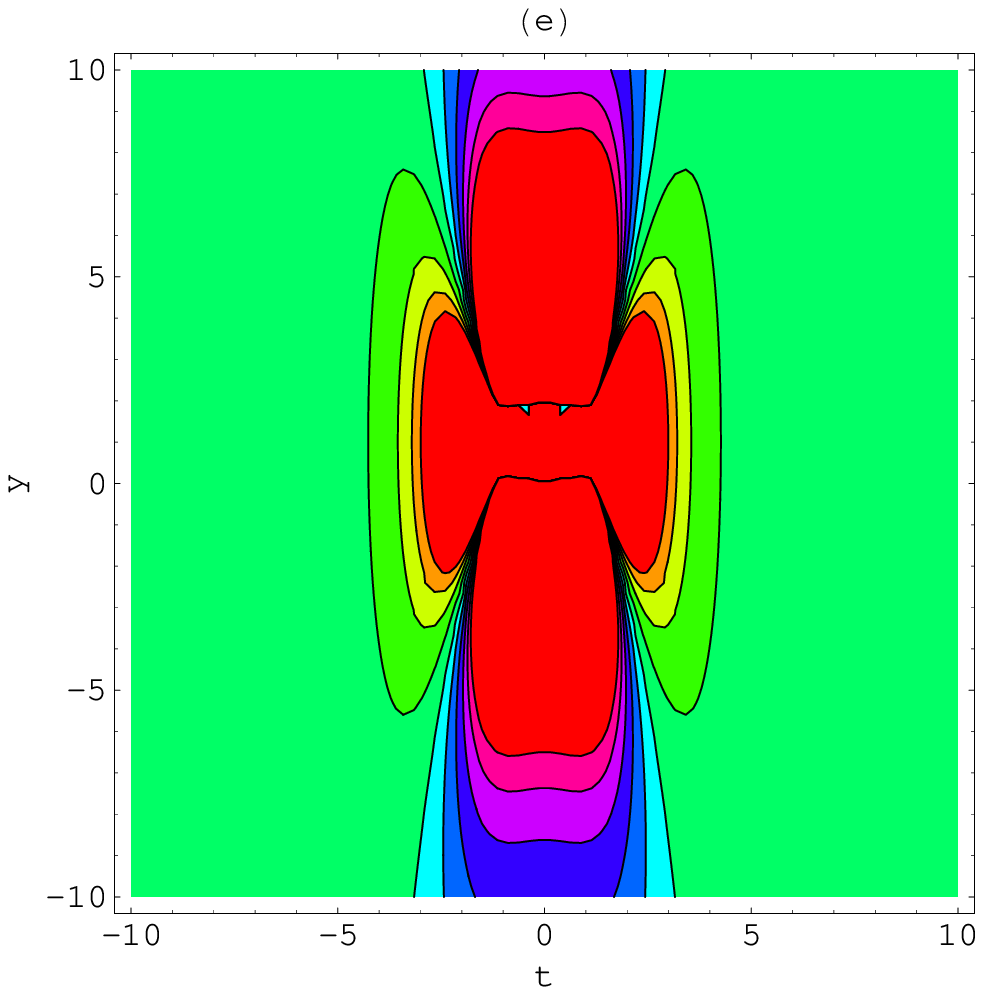}
\includegraphics[scale=0.35,bb=-290 620 10 10]{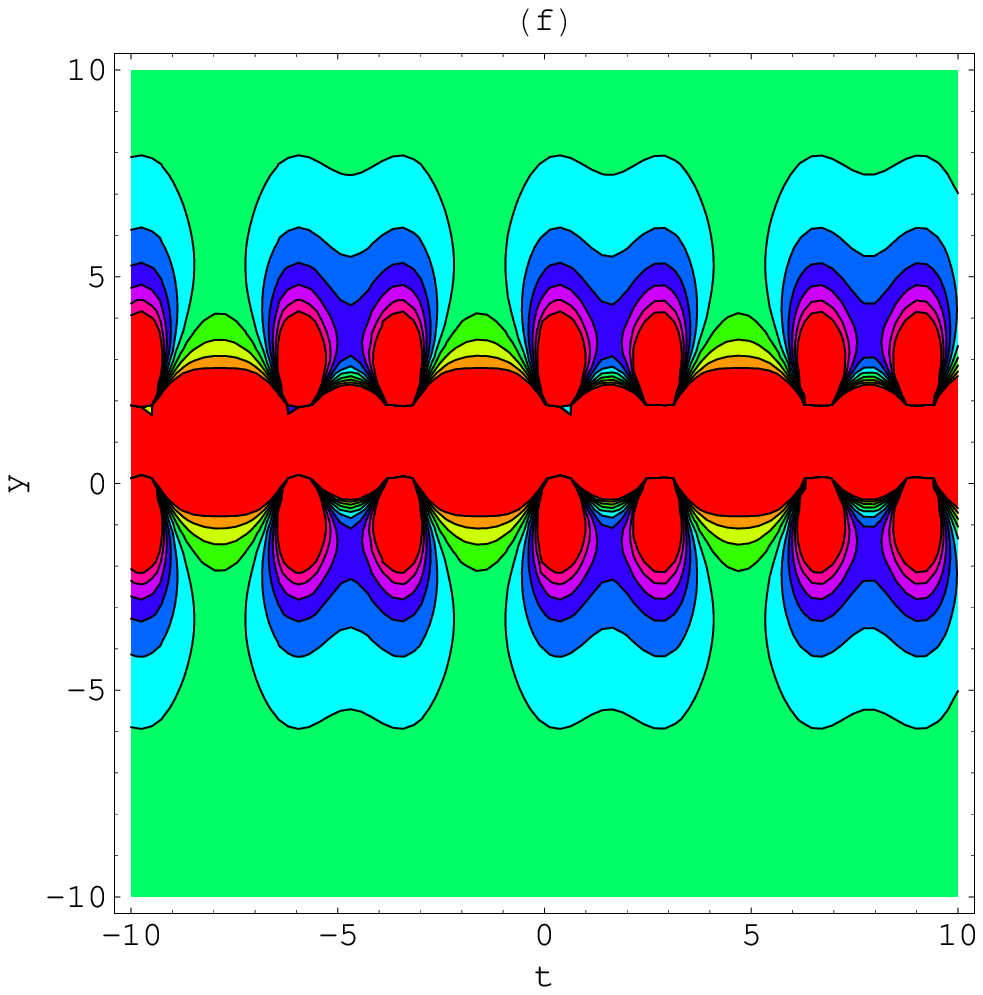}
\vspace{7.5cm}
\begin{tabbing}
\textbf{Fig. 11}. Rogue wave (17) with $\mu=\nu=\Theta _0=1$, $x=0$,
$\zeta_0=-1$, $\zeta_1=2$,\\ $\beta(t)=1$ in (a) (d), $\beta(t)=t$
in (b) (e) and $\beta(t)=\cos t$ in (c) (f).\\
\end{tabbing}

\quad In addition to this (2+1)-dimensional vcKP equation, this
vcsca can also be applied to the (3+1)-dimensional generalized KP
equation with variable coefficients [19], the generalized (3 +
1)-dimensional variable-coefficient nonlinear-wave equation [20]
based on the symbolic computation [21-26].\\

\noindent {\bf Compliance with ethical standards}\\

\quad {\bf Conflict of interests} The authors declare that there is
no conflict of interests regarding the publication of this article.

\quad {\bf Ethical standard} The authors state that this research
complies with ethical standards. This research does not involve
either human participants or animals.


\begin{thebibliography}{}
%
%
\bibitem{Ref1}
Wang, Y.Y.,  Zhang, J.F.:  Variable-coefficient KP equation and
solitonic solution for two-temperature ions in dusty plasma. Phys.
Lett. A., 352(1), 155-162 (2006).
\bibitem{s1}   Yao, Z.Z.,  Zhang, C.Y., et al.:  Wronskian and grammian determinant solutions for a
variable-coefficient Kadomtsev-Petviashvili equation. Commun. Theor.
Phys., 49(5), 1125-1128 (2008).
\bibitem{s1}  Wu, J.P.:  Bilinear B\"{a}cklund transformation for a variable-coefficient Kadomtsev-Petviashvili
 equation. Chin. Phys.s Lett., 28(6), 060207 (2011).
\bibitem{s2} Liu, J.G., Zhu,W.H., Zhou, L.: Breather wave solutions for the Kadomtsev-Petviashvili
equation with variable coefficients in a fluid based on the
variable-coefficient three-wave approach. Math. Method. Appl. Sci.,
DOI: 10.1002/mma.5899 (2019).
\bibitem{s3} Jia, X.Y., Tian, B., Du, Z., Sun, Y.,  Liu, L.:  Lump and rogue waves for the variable-coefficient
Kadomtsev-Petviashvili equation in a fluid. Mod. Phys. Lett. B.,
32(10), 1850086 (2018).
\bibitem{s4} Liu, J.G., Zhu,W.H., Zhou, L.: Interaction Solutions for Kadomtsev-Petviashvili Equation with Variable
Coefficients. Commun. Theor. Phys., 71,  793-797 (2019).
\bibitem{s1}  Grimshaw, R.,  Pelinovsky, E.,  Taipova, T.,  Sergeeva,  A.: Rogue internal waves in the ocean: Long wave model. Eur. Phys.
J. Spec. Top., 185,  195-208 (2010).
\bibitem{s5}  Zuo, D.W.,  Gao, Y.T.,  Xue, L.,  Feng, Y.J.,  Sun, Y.H.: Rogue waves for the generalized nonlinear
Schr\"{o}dinger-Maxwell- Bloch system in optical-fiber
communication. Appl. Math. Lett., 40,  78-83 (2015).
\bibitem{s6}  He, J.S.,  Charalampidis, E.G.,  Kevrekidis, P.G.,  Frantzeskakis, D.J.: Rogue waves in nonlinear Schr\"{o}dinger models with
variable coefficients: Application to Bose-Einstein condensates.
Phys. Lett. A, 378(5-6),  577-583 (2014).
\bibitem{s7} Li, B.Q., Ma, Y.L.: Rogue waves for the optical fiber system with variable
coefficients. Optik, 158, 177-184 (2018).
\bibitem{s11}  Ankiewicz, A.,  Akhmediev,  N.: Rogue wave-type solutions of the mKdV equation and their relation to known NLSE rogue wave
solutions. Nonlinear Dyn., 91(3), 1931-1938 (2018).
\bibitem{s12} Su, J.J., Gao, Y.T., Ding, C.C.: Darboux transformations and rogue wave solutions of a generalized AB system for the geophysical
flows. Appl. Math. Lett., 88, 201-208 (2019).
\bibitem{s13} Wang, X.B., Zhang, T.T., Dong, M.J.: Dynamics of the breathers and rogue waves in the higher-order nonlinear Schr\"{o}dinger
equation. Appl. Math. Lett., 86, 298-304 (2018).
\bibitem{s14} Clarkson, P.A.,  Dowie, E.: Rational solutions of the Boussinesq
equation and applications to rogue waves. Trans. Math. Appl., 1(1),
 tnx003 (2017).
\bibitem{s15} Zha,Q.L.: A symbolic computation approach to constructing rogue waves with a controllable center in the nonlinear
systems. Comput. Math. Appl., 75(9),  3331-3342 (2018).
\bibitem{s16} Liu, W.H., Zhang, Y.F.:  Multiple rogue wave solutions of the (3+1)-dimensional
Kadomtsev-Petviashvili-Boussinesq equation. Z. Angew. Math. Phys.,
 70, 112 (2019).
\bibitem{s17} Zhao, Z.L., He, L.C., Gao, Y.B.: Rogue Wave and Multiple Lump Solutions of
the (2+1)-Dimensional Benjamin-Ono Equation in Fluid Mechanics.
Complexity,, 2019, 8249635 (2019).
\bibitem{s18} Liu, W.H., Zhang, Y.F.: Multiple rogue wave solutions for a (3+1)-dimensional Hirota
bilinear equation. Appl. Math. Lett., 98,  184-190 (2019).
\bibitem{s19} Liu, J.G., Ye, Q.: Stripe solitons and lump solutions for a generalized
Kadomtsev-Petviashvili equation with variable coefficients in fluid
mechanics. Nonlinear Dyn., 96(1), 23-29 (2019)
\bibitem{s19} Deng, G.F.,  Gao, Y.T.: Integrability, solitons, periodic and travelling waves of a generalized
(3+1)-dimensional variable-coefficient nonlinear-wave equation in
liquid with gas bubbles. Eur. Phys. J. Plus., 132(6), 255-271
(2017)
\bibitem{s19} Gaillard, P.: Rational solutions to the KPI equation and multi rogue waves. Ann. Phys., 367,  1-5 (2016).
\bibitem{s20} Yin, Y.H., Ma, W.X., Liu, J.G., L\"{u}, X., Diversity of exact solutions to a (3+1)-dimensional
 nonlinear evolution equation and its reduction, Comput. Math. Appl., 76, 1275-1283 (2018).
\bibitem{s21} L\"{u}, X., Lin, F.H., Qi, F.H.: Analytical study on a two-dimensional Korteweg-de Vries model with bilinear
representation, B\"{a}cklund transformation and soliton solutions.
Appl. Math. Model., 39, 3221-3226 (2015)
\bibitem{s22} Xu, G.Q.,  Wazwaz, A.M.: Characteristics of integrability, bidirectional solitons and
localized solutions for a (3 + 1)-dimensional generalized breaking
soliton equation. Nonlinear Dyn.,  96, 1989-2000 (2019).
\bibitem{s23}  Li, Y.Z., Liu, J.G.: New periodic solitary wave solutions for the
new (2+1)-dimensional Korteweg-de Vries equation. Nonlinear Dyn.,
91(1), 497-504 (2018).
\bibitem{s24} Lan, Z.Z., Su, J.J.: Solitary and rogue waves with controllable backgrounds for
the non-autonomous generalized AB system. Nonlinear Dyn., 96,
2535-2546 (2019)


\end{thebibliography}


\end{document}